\documentclass[lettersize,journal]{IEEEtran}

\usepackage[colorlinks=true, linkcolor=black, citecolor=black, urlcolor=black]{hyperref}
\usepackage{amssymb}
\usepackage{tikz}
\usepackage{xspace}
\newcommand{\sysname}{\textsc{NanoZone}\xspace}
\usepackage{cleveref}
\crefformat{section}{#2§#1#3}
\crefname{section}{§}{§§}
\usepackage{pifont}
\usepackage{adjustbox}
\usepackage{multirow}
\usepackage{enumitem}
\usepackage{threeparttable}
\newcommand{\RNum}[1]{\uppercase\expandafter{\romannumeral #1\relax}}
\usepackage[misc]{ifsym}
\usepackage{booktabs}
\usepackage{tabularx}
\usepackage{array}
\usepackage{xurl}
\usepackage[figure]{hypcap}
\begin{document}

\title{\sysname: Scalable, Efficient, and Secure Memory Protection on Arm CCA}


\author{Shiqi~Liu,
    Yongpeng~Gao,
    Mingyang~Zhang,
    and Jie~Wang
    \thanks{This work was supported in part by the National Natural Science Foundation of China under Grant 62202194 and in part by the CCF-Ant Research Fund. \emph{(Corresponding author: Jie~Wang.)}}

    \thanks{Shiqi~Liu, Yongpeng~Gao, Mingyang~Zhang, and Jie~Wang are with Hubei Key Laboratory of Distributed System Security, Hubei Engineering Research Center on Big Data Security, School of Cyber Science and Engineering, Huazhong University of Science and Technology, Wuhan, 430074, China (e-mail: shiqiliu@hust.edu.cn; sternen\_hust@hust.edu.cn; zoneshiyi@hust.edu.cn; wangjie\_s@hust.edu.cn).}
}

\markboth{IEEE Transactions on Dependable and Secure Computing}
{LIU \MakeLowercase{\textit{et al.}}: \sysname: Scalable, Efficient, and Secure Memory Protection on Arm CCA}


\maketitle

\begin{abstract}
    Arm Confidential Compute Architecture (CCA) provides hardware primitives for confidential computing, but its standard CVM-based deployment model still isolates software only at the VM granularity. This leaves two critical gaps: the guest OS remains in the Trusted Computing Base, and applications lack any intra-process isolation boundary, exposing secrets to vulnerabilities such as Heartbleed. While intra-address-space isolation offers a theoretical solution, realizing it under an adversarial OS faces a fundamental scalability-efficiency trade-off: existing hardware primitives are either strictly limited in domain capacity or incur prohibitive privileged-switching latency.

    To bridge this gap, we propose \sysname, a novel isolation architecture built on Arm CCA primitives. Rather than extending the realm-world software stack, \sysname protects normal-world processes and enforces fine-grained intra-process isolation from a minimal root-world monitor. \sysname unifies the speed of user-space permission switching with the capacity of physical address space isolation to achieve effectively unlimited domain scalability. To mask the latency of privileged transitions, it employs a locality-aware scheduling policy that maximizes execution residency within the fast user-level tier. To defeat an adversarial OS, \sysname anchors its root of trust in the root world, offloading critical isolation enforcement away from the untrusted kernel. Against intra-process adversaries, it further leverages hardware-assisted Code-Pointer Integrity (CPI) to prevent domain-switching abuse. We prototyped \sysname on Arm's official emulator and on physical hardware. Evaluation on real-world server applications shows that, compared to process-level isolation systems, our fine-grained protection preserves about 95\% of their throughput.
\end{abstract}


\section{Introduction}
Confidential Computing protects tenant assets using hardware-based Trusted Execution Environments (TEEs) such as AMD SEV~\cite{amd2023sev} and Intel TDX~\cite{intel2023tdx}. In the Arm ecosystem, the recently introduced \emph{Confidential Compute Architecture (CCA)}~\cite{arm2023cca} adds a \emph{realm world} to Armv9-A to host Confidential Virtual Machines (CVMs).
However, this coarse-grained isolation model leaves CCA exposed to two critical vectors. 
(1) \textbf{The Threat of an Adversarial Kernel.} Current architectures implicitly trust the guest OS, forcing it into the Trusted Computing Base (TCB). With monolithic kernels (e.g., Linux) exceeding 40M lines of code (LoCs), this design creates a massive attack surface where privilege-escalation bugs~\cite{zeng2023retspill,lin2022dirtycred} allow an adversary to compromise the entire CVM. 
(2) \textbf{Unmitigated Intra-Process Risks.} While CCA shields the VM from the host, it lacks primitives for fine-grained protection within the application itself. Complex software like OpenSSL (900K+ LoCs) handles secrets in only a fraction of its code base. Without intra-address-space boundaries, a single memory safety bug like Heartbleed~\cite{durumeric2014matter} can leak secrets from non-critical code paths.

In response to the threat of an adversarial kernel, prior research has focused on TCB reduction. Gramine-TDX~\cite{kuvaiskii2024gramine} replaces the monolithic kernel with a leaner LibOS ($\sim$57K LoCs), while RContainer~\cite{zhourcontainer} minimizes the trusted computing base via a mini-OS ($<$5K LoCs) on the CCA platform. Shelter~\cite{zhang2023shelter} goes further by completely removing the OS from the trust boundary, enforcing process-level isolation. While Shelter effectively shields the application from the kernel, it fails to mitigate the \emph{internal attack surface} described above.

Intra-address-space isolation is not a new concept in TEEs. For instance, in the context of Intel SGX~\cite{intel2023sgx}, Nested Enclave~\cite{park2020nested} partitions an enclave into outer and inner layers through hardware modifications. However, the switching time between these layers (1.1 \textmu s, comparable to native ecall/ocall latency) makes it impractical for high-frequency domain transitions. To further optimize performance, LightEnclave~\cite{gu2022hardware} employs Memory Protection Keys (MPK)~\cite{intel2024mpk} to partition an enclave into multiple domains. The performance gain stems from complete user-space domain switches that require only about 20 cycles. However, applying these established concepts to Arm CCA presents unique challenges. The CVM-style isolation in CCA and enclave-style isolation in SGX differ substantially in their architectural abstractions and hardware primitives. Furthermore, although solutions like NestedSGX~\cite{wang2024road}, Veil~\cite{ahmad2023veil}, and CAP-VMs~\cite{sartakov2022cap} have demonstrated intra-CVM isolation, adapting them to the standard Armv9 architecture is impractical. These approaches rely on either experimental hardware extensions (e.g., CHERI capabilities~\cite{watson2015cheri} in CAP-VMs) or vendor-specific features (e.g., AMD VMPLs in Veil) that are absent in the Arm CCA specification. Consequently, porting them to Arm would incur prohibitive overheads from complex hypervisor emulation.

For multi-tenant servers that hold secrets for mutually distrusting clients within a single process, such as per-client TLS keys in Nginx, these limitations call for a deployable Arm CCA design that combines intra-process isolation with a root of trust outside the OS.

This paper introduces \sysname, a hardware-assisted intra-process isolation system built on Arm CCA primitives. \sysname aims to provide fine-grained isolation under an adversarial OS while minimizing the trusted software stack. To this end, \sysname places protected processes in the normal world and anchors security enforcement in a minimal root-world monitor at EL3, rather than extending the Realm/CVM software stack and trusting the additional Realm Management Monitor (RMM, 46K LoCs) at Realm EL2.

Efficient fine-grained isolation also requires fast domain switching. Intel MPK is a promising model because it supports user-space permission changes without costly kernel traps or TLB flushes, but it is unavailable on Arm. Although Armv7 previously provided a similar feature through memory domains~\cite{chen2016shreds,zhou2014armlock}, its Domain Access Control Register is inaccessible from user space, and the feature has been deprecated in AArch64. Fortunately, Arm recently introduced the \emph{Permission Overlay Extension (POE)}~\cite{arm2024pio} in Armv8.9-A, making efficient user-space domain switching feasible on Arm. \sysname partitions memory pages into isolation domains by encoding POE indices, analogous to MPK protection keys, in Page Table Entries (PTEs). On authorized domain switches, \sysname temporarily grants access to sensitive memory by updating POE permission registers in user space, keeping the access window short. Because POE permissions are maintained per core, other cores remain restricted by their own permission-register state.

While the high-level design of \sysname may seem straightforward, it faces challenges spanning scalability, efficiency, and security. The core contribution of this paper lies in balancing the trade-offs across these three dimensions.

\noindent \textbf{Scalability Challenge: Expanding Domains beyond Hardware Limits.}
POE is strictly limited to 7 usable domains, which is insufficient for multi-tenant server workloads. While Arm CCA's GPT mechanism supports unlimited domains, relying solely on it necessitates frequent privileged interventions, which undermines the lightweight nature of user-space isolation~\cite{liu2025more}. The core challenge is to architect a mechanism that achieves infinite scalability within the rigid constraints of hardware resources.
\sysname overcomes this by integrating complementary hardware features into a cohesive three-tier zone design (\cref{subsec:extend}). First, it combines POE with the \emph{Permission Indirection Extension (PIE)}~\cite{arm2024pio}, mapping unused PIE indexes to existing POE indexes to expand user-level domains from 7 to 28. Second, to break the hard limit, it exploits \emph{Granule Protection Tables (GPT)}~\cite{arm2023introcca} and \emph{GPC3} bypass windows~\cite{bypass2024arm} to reuse these indexes across distinct Physical Address Spaces (PAS). This hierarchical approach enables it to scale to an \emph{unlimited} number of domains.

\begin{figure}[!t]
  \centering
  \includegraphics[width=\linewidth]{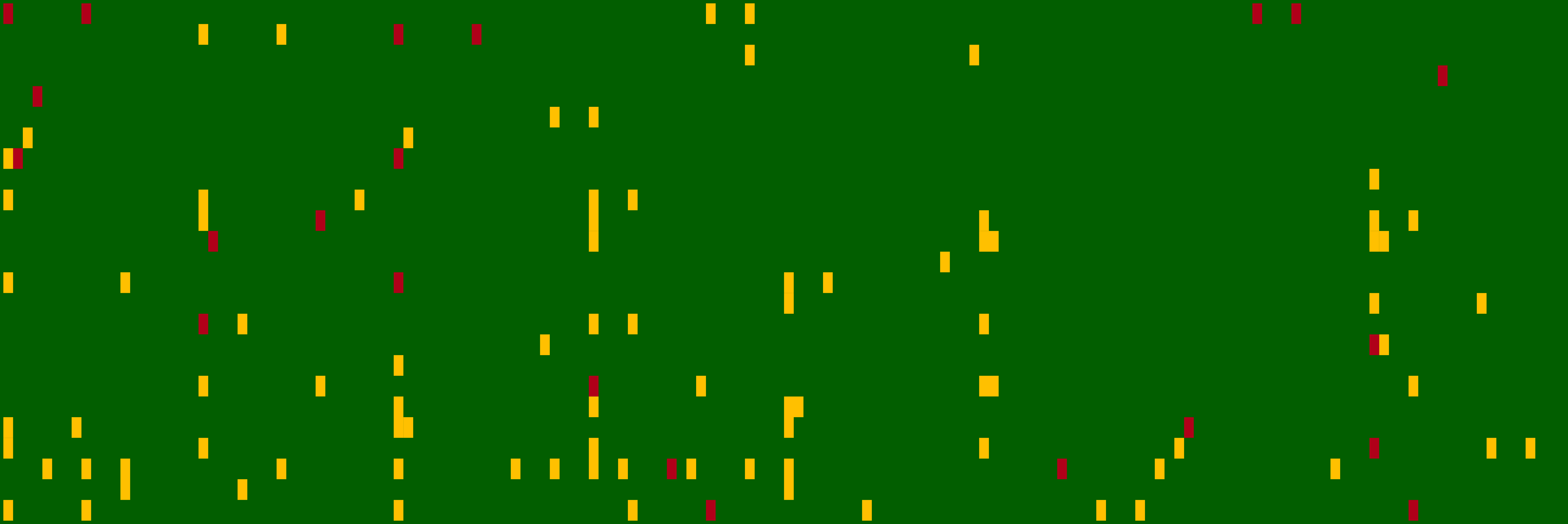}
  \caption{Domain-switching scenarios for 1,000 Memcached requests in \sysname (green: POE switches; yellow: PIE switches; red: PAS switches).}
  \label{fig:heatmap}
\end{figure}

\noindent \textbf{Efficiency Challenge: Amortizing the Cost of Hierarchical Isolation.}
The hierarchical scalability established above introduces a fundamental trade-off: accessing deeper tiers (PIE or PAS) incurs prohibitively higher switching costs. Unlike lightweight POE switches, crossing these boundaries requires trapping into higher privilege levels, resulting in $\sim$80$\times$ higher latency. The challenge is to sustain high throughput when the underlying isolation mechanism is inherently expensive.
\sysname addresses this by enforcing a tier-aware execution model that accounts for the heterogeneous costs of domain switching. We exploit request locality to mask the latency of privileged traps, strategically scheduling requests to maximize residency within the currently active low-latency POE layer. As shown in \autoref{fig:heatmap}, this approach maintains a 96.72\% hit rate in the fastest tier, effectively restricting expensive transitions to the bare minimum. Consequently, \sysname amortizes the heavy hardware isolation costs, reducing the average domain-switch latency to near-native speeds (only 4.87\% of a raw privileged switch).

\noindent \textbf{Security Challenge: Enforcing Integrity against an Adversarial Kernel.}
Standard Control-Flow Integrity (CFI) relies on the critical premise that the OS enforces memory isolation to prevent code and pointer corruption. In our threat model, however, this assumption is invalidated. An adversarial OS can subvert isolation boundaries by manipulating page tables, rendering user-space CFI mechanisms ineffective. The challenge, therefore, is to restore the integrity of the isolation layer as a prerequisite for robust CFI. \sysname addresses this dependency by re-anchoring the root of trust in the monitor. First, to restore isolation, we incorporate a security module (\cref{subsec:reduction}) that validates page table updates and protects sensitive registers, precluding the OS from remapping or tampering with domain memory. With isolation secured, we then enforce CFI by combining the \emph{Guarded Control Stack (GCS)}~\cite{arm2023manual} with a monitor-managed \emph{Pointer-Integrity Memory (PIM)} (\cref{subsec:cfi}), ensuring execution flow remains tamper-resistant even against kernel-level interference.

Since no commercial hardware yet supports the required Arm features, we built two prototypes: a functional version on Arm's official Fixed Virtual Platform and a performance version on a hardware SoC. Security evaluations confirm that \sysname effectively mitigates both intra-process vulnerabilities and privileged attacks via surface analysis (\cref{subsec:analysis}) and CVE reproduction (\cref{subsec:cve}).
Performance assessments on micro-benchmarks (\cref{subsec:microbenchmarks}) and real-world applications (\cref{subsec:real_world}) show that \sysname incurs only moderate overhead. 
For instance, while Nginx exhibits a 21.84\% average throughput overhead, the marginal cost of our fine-grained protection remains small: compared to the state-of-the-art Shelter~\cite{zhang2023shelter} (which lacks intra-process isolation), \sysname introduces only a 5.15\% relative drop.

\begin{table*}[!t]
\centering
\caption{Acronyms and key registers used in this paper.}
\label{tab:glossary}
\small
\begin{adjustbox}{max width=\linewidth}
\begin{tabular}{ll|ll|ll}
\toprule
\textbf{Abbrev.} & \textbf{Full Name} &
\textbf{Abbrev.} & \textbf{Full Name} &
\textbf{Register} & \textbf{Full Name} \\
\midrule
CCA  & Confidential Compute Architecture    & GPC  & Granule Protection Check        & \texttt{POR\_EL0}    & POE permission register \\
RME  & Realm Management Extension           & GPC3 & Granule Protection Check 3      & \texttt{PIRE0\_EL1}  & PIE permission register \\
CVM  & Confidential Virtual Machine         & GPF  & Granule Protection Fault        & \texttt{GPCBW\_EL3}  & GPC bypass-window register \\
PAS  & Physical Address Space               & GCS  & Guarded Control Stack           & \texttt{GCSPR\_EL0}  & GCS pointer register \\
GPT  & Granule Protection Table             & CPI  & Code-Pointer Integrity          & \texttt{TPIDRRO\_EL0}& Read-only thread pointer register \\
POE  & Permission Overlay Extension         & PIM  & Pointer-Integrity Memory        & & \\
PIE  & Permission Indirection Extension     & PTE  & Page Table Entry                & & \\
\bottomrule
\end{tabular}
\end{adjustbox}
\end{table*}

In summary, we make the following contributions:
\begin{itemize}
    \item We propose a hierarchical model to resolve the \emph{scalability-efficiency trade-off} in Arm hardware. It unifies the speed of POE with the scale of GPT, enabling unlimited isolation domains without sacrificing user-space performance.
    
    \item We redesign protection mechanisms for the threat model where the OS is adversarial. Anchoring trust in the monitor, \sysname enforces control-flow integrity (via GCS and CPI) independent of kernel-managed page tables.
    
    \item We show that our \emph{locality-aware scheduling} effectively masks the latency of privileged transitions. Evaluation confirms that it retains about 95\% of the throughput of process-level isolation, proving the practicality of fine-grained isolation.
\end{itemize}

\section{Background and Motivation} \label{sec:background}

\subsection{Hardware Background}
For ease of reference, \autoref{tab:glossary} summarizes the main architectural acronyms and key registers used in the paper.

\begin{figure}[!t]
  \centering
  \includegraphics[width=\linewidth]{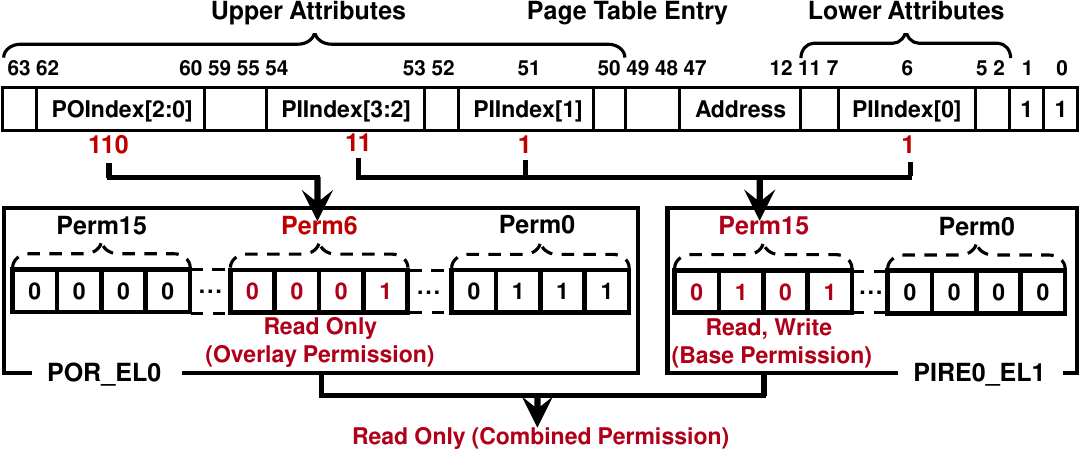}
  \caption{Combination of permission indirection and overlay.}
  \label{fig:pio}
\end{figure}

\noindent \textbf{Permission Indirection Extension (PIE).} Arm previously used a direct memory permission model, where permissions for each memory page were determined by the permission bits in the PTE. In this model, modifying the permissions of a domain requires updating the PTEs of all pages associated, making it inefficient for domain-based isolation. Additionally, the limited number of PTE bits restricts scalability for supporting new permissions. To address these limitations, Arm introduced an indirect permission model~\cite{arm2024pio} in Armv8.9-A. This model manages user-space permissions using the 64-bit \texttt{PIRE0\_EL1} register, where each 4 bits specify the permissions for a domain, supporting up to 16 domains. The original PTE permission bits (i.e., bit 54 \texttt{UXN}, bit 53 \texttt{PXN}, bit 51 \texttt{WRITE/DBM}, and bit 6 \texttt{USER/AP[1]}) are repurposed as indexes to indicate a page's domain. Notably, the direct and indirect permission models are compatible. As shown in \autoref{fig:pio}, in the direct model, a user-space memory page with read-write permissions has the bits \texttt{UXN} = 1, \texttt{PXN} = 1, \texttt{WRITE} = 1, and \texttt{USER} = 1 set. In the indirect model, these bits instead serve as an index in \texttt{PIRE0\_EL1} (i.e., \texttt{perm15}), which is pre-configured with read-write permissions (i.e., \texttt{0b0101}). Among the 16 domains supported by \texttt{PIRE0\_EL1}, 12 are pre-configured with fixed permissions to cover common use cases. 

\noindent \textbf{Permission Overlay Extension (POE).}
Although PIE can configure base permissions, permission switching requires trapping to privileged mode and flushing the TLB, leading to inefficiencies. To address this, Arm also introduced the POE~\cite{arm2024pio} in Armv8.9-A, delegating part of permission management to user-space. As shown in \autoref{fig:pio}, base permissions can be further restricted by overlay permissions stored in the 64-bit \texttt{POR\_EL0} register, which do not need to be cached in the TLB. During runtime, the MMU transparently combines and checks both base and overlay permissions. POE uses three unused bits [62:60] in the PTE as the index, supporting up to 8 POE domains\footnotemark. However, the POE index defaults to 0, where \texttt{perm0} is fixed with full permissions (i.e., \texttt{0b0111}, read, write, and execute) to ensure that default base permissions are not overridden. As a result, only 7 POE domains are available for general use.

\footnotetext{In Armv9.4-A, the 128-bit translation tables~\cite{arm2023manual} extend support for a 4-bit POE index, enabling all 16 POE domains. However, software stack support for the 128-bit translation tables is still under development.}

\noindent \textbf{Guarded Control Stack (GCS).}
Arm introduced GCS~\cite{arm2023manual} in Armv9.4-A to defend against ROP~\cite{shacham2007geometry}, similar to the x86 Shadow Stack~\cite{intel2023shadow}. GCS leverages a new permission (i.e., \texttt{perm11} in \texttt{PIRE0\_EL1}) introduced by PIE, allowing non-privileged read operations while disallowing writes, except through the specific \texttt{GCSSTR} instruction. Notably, POE does not apply to GCS. When enabled, GCS allocates a protected stack for each thread in PIE \texttt{domain11}, pointed to by the GCS pointer register (\texttt{GCSPR\_EL0}). During branch instructions with a link (e.g., \texttt{BL}, \texttt{BLR}), the CPU saves the return address in the link register (\texttt{LR}) and transparently pushes it onto the protected stack. Similarly, during a \texttt{RET} instruction, the CPU verifies the \texttt{LR} against the top of the protected stack to ensure return address integrity.

\begin{figure}[!t]
  \centering
  \includegraphics[width=\linewidth]{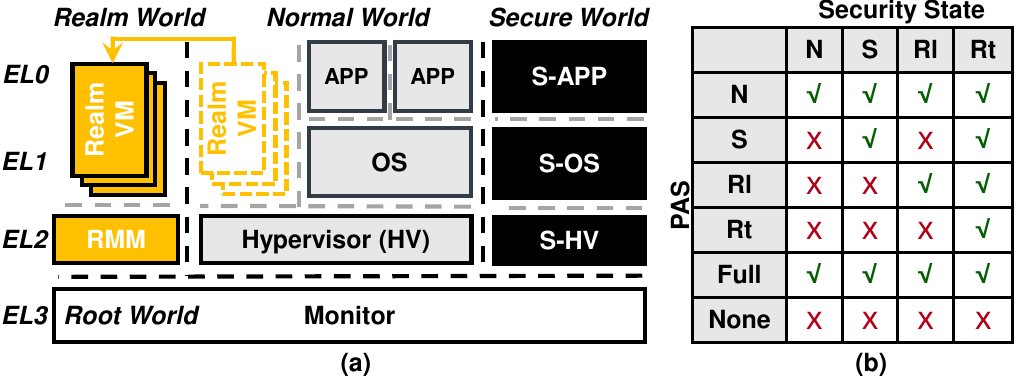}
  \caption{The architecture of CCA and its PAS access permissions.}
  \label{fig:cca}
\end{figure}

\noindent \textbf{Confidential Compute Architecture (CCA).}
In the Armv9.2-A, Arm introduced the Confidential Compute Architecture (CCA)~\cite{arm2023cca} with a hardware primitive called the Realm Management Extension (RME)~\cite{arm2023introcca}. CCA introduces a new execution environment called the realm world and separates the entire Exception Level 3 (EL3) into a distinct root world. Confidential Virtual Machines (CVMs), originally created by the normal-world hypervisor, are executed in the realm world, as shown in \autoref{fig:cca}(a). 
RME achieves fine-grained physical memory access control by leveraging two main steps: configuring Physical Address Spaces (PAS) through the Granule Protection Table (GPT) and enforcing PAS access checks with the Granule Protection Check (GPC). Specifically, the GPT maps each physical memory page (at a 4KB granularity) to one of several PAS: \texttt{normal}, \texttt{secure}, \texttt{realm}, \texttt{root}, \texttt{full-access}, or \texttt{no-access}. During memory accesses, the MMU consults the GPT to retrieve PAS of the target memory page and performs a GPC to validate access permissions based on the current security state, as illustrated in \autoref{fig:cca}(b). If an access is invalid, a Granule Protection Fault (GPF) is triggered immediately. The most privileged root world ensures that other worlds cannot modify the GPT or GPC's registers.

\noindent \textbf{Granule Protection Check 3 (GPC3).}
Armv9.6-A introduced GPC3, an upgraded GPC that lets designated memory regions bypass the check by configuring the GPC bypass window register (\texttt{GPCBW\_EL3})~\cite{bypass2024arm}. Bypass windows are configurable from 1GB to 64GB. 
Each core has its own \texttt{GPCBW\_EL3}. A core can directly access its configured bypass windows, while other cores must undergo GPC.

\subsection{Limitations of Existing Solutions} \label{subsec:motivation}

Intra-process isolation is the leading technique for enforcing the principle of least privilege. Existing mechanisms (see \autoref{tab:comparisons}) partition memory into distinct domains within a single address space and use specialized hardware instructions to switch between them. For example, lwC~\cite{litton2016light} builds contexts by assigning each a separate virtual memory mapping via page tables. Context switches require updating the page-table base, and even with PCID support to avoid global TLB flushes, the transition still traps into privileged mode, incurring non-negligible overhead.

Unlike page-table-based approaches, Intel MPK~\cite{intel2024mpk} enables unprivileged domain switching, achieving $<2\%$ overhead when isolating SSL keys in Nginx~\cite{voulimeneas2022you}. 
However, MPK supports only 16 domains, which hampers scalability in multi-tenant workloads, prompting new designs that lift this limit. 
For instance, Donky~\cite{schrammel2020donky} leverages the 10 reserved bits of the PTE to support up to 1,024 domains but requires substantial hardware modifications. 
libmpk~\cite{park2019libmpk} employs an LRU-based algorithm to evict inactive domains, reusing their domain IDs without hardware changes. 
However, evicting a domain involves modifying the PTEs of all associated memory pages. 
As the number of pages or domains grows, eviction costs and rates increase, leading to significant overhead. 
EPK~\cite{gu2022epk} leverages Intel's Extended Page Table (EPT) to reuse MPK domain IDs. 
Since the \texttt{VMFUNC} instruction used for EPT switching is also non-privileged, domain switching can be efficiently performed in user space. 
LightZone~\cite{yuan2024lightzone}, like lwC, creates separate memory mappings for domains to enhance scalability, theoretically supporting up to 65,536 domains. 
The key difference is that LightZone runs processes in kernel space to avoid trapping overhead of domain switching. 
Crucially, however, both EPK and LightZone depend on a trusted OS or hypervisor to manage these isolation structures. This dependency not only inflates the TCB but also renders them insecure under the adversarial kernel threat model addressed by \sysname.

\begin{table}[t]
  \caption{Comparison of intra-process/enclave isolation.}
  \label{tab:comparisons}
  \centering
  \small
  \begin{adjustbox}{max width=\linewidth}
  \begin{tabular}{@{}lccccc@{}}
    \toprule
    \multirow{2}{*}{\textbf{System}} &
    \multirow{2}{*}{\textbf{Efficiency}\textsuperscript{1}} &
    \multirow{2}{*}{\textbf{Scalability}\textsuperscript{2}} &
    \multicolumn{2}{c}{\textbf{Security}\textsuperscript{3}} &
    \multirow{2}{*}{\textbf{Platform}} \\
    \cmidrule(lr){4-5}
    & & & PA & SA & \\
    \midrule
    lwC~\cite{litton2016light}              & Slow & Many   & \ding{55} & \ding{55} & Portable \\
    MPK~\cite{intel2024mpk}                 & Fast & 16     & \ding{55} & \ding{55} & Intel    \\
    Donky~\cite{schrammel2020donky}         & Fast & 1,024  & \ding{55} & $\dagger$    & RISC-V   \\
    libmpk~\cite{park2019libmpk}            & Slow & Many   & \ding{55} & \ding{55} & Intel    \\
    EPK~\cite{gu2022epk}                    & Fast & 7,680  & \ding{55} & $\dagger$    & Intel    \\
    LightZone~\cite{yuan2024lightzone}      & Fast & 65,536 & \ding{55} & $\dagger$    & Arm      \\
    Nested Enclave~\cite{park2020nested}    & Slow & Many   & \ding{51} & $\dagger$    & Intel    \\
    LightEnclave~\cite{gu2022hardware}      & Fast & 16     & \ding{51} & $\dagger$    & Intel    \\
    HiveTEE~\cite{huang2026hivetee}         & Slow & Many   & \ding{51} & $\dagger$    & Arm      \\
    \midrule
    \textbf{\sysname} & \textbf{Fast}\textsuperscript{4} & \textbf{Many} & \textbf{\ding{51}} & \textbf{\ding{51}} & \textbf{Arm} \\
    \bottomrule
  \end{tabular}
  \end{adjustbox}

  \vspace{2pt}
  \begin{minipage}{0.97\linewidth}
    \footnotesize
    \textsuperscript{1}\textbf{Efficiency} refers to domain-switching performance. We classify efficiency as “Fast” if domain switching can be performed primarily in user space (e.g., in the low hundreds of cycles).\\
    \textsuperscript{2}\textbf{Scalability} refers to the maximum number of supported domains.\\
    \textsuperscript{3}\textbf{Security} refers to the ability to resist the PA (\underline{P}rivilege \underline{A}ttack) from the OS or the internal SA (\underline{S}witch \underline{A}buse). $\dagger$\ indicates incomplete defense.\\
    \textsuperscript{4}Due to \sysname's hierarchical design, we consider the amortized switching cost here.
  \end{minipage}
\end{table}

While intra-enclave isolation effectively mitigates threats posed by privileged adversaries, it struggles with efficiency and scalability. Nested Enclave~\cite{park2020nested} incurs substantial overhead during switches between outer and inner enclaves, whereas MPK-based solutions such as LightEnclave~\cite{gu2022hardware} inherently restrict the number of isolation domains to 16. The concurrent HiveTEE~\cite{huang2026hivetee} lifts this limit on Arm CCA by reusing MTE tags across GPTs, but every entry into its domains still traps to the Monitor. On the other hand, domain-switching abuse creates internal attack vectors that are often overlooked. 
lwC~\cite{litton2016light} and libmpk~\cite{park2019libmpk} ignore these threats entirely. 
More advanced designs like Donky~\cite{schrammel2020donky}, EPK~\cite{gu2022epk}, LightEnclave~\cite{gu2022hardware}, Nested Enclave~\cite{park2020nested}, LightZone~\cite{yuan2024lightzone}, and HiveTEE~\cite{huang2026hivetee} implement \emph{gate-centric defenses}. 
By authenticating the caller's identity, execution state, or entry target, they effectively prevent \textbf{unauthorized entry} (arbitrary jumps or returns to the trampoline entry). 
Furthermore, they mitigate \textbf{mid-trampoline injection} either through hardware enforcement (Donky, Nested Enclave) or by verifying post-switch execution states, such as random nonces (EPK, LightEnclave) or call-site allowlists (LightZone). 
However, these defenses remain fundamentally vulnerable to \textbf{borrowed-gate attacks} (or confused-deputy attacks). 
In this scenario, an intra-process adversary exploits memory corruption to overwrite control-flow data (e.g., return addresses or function pointers) in untrusted code, redirecting execution to a \textit{legitimate} call site that invokes the trampoline. 
Since the domain switch is triggered from a valid location, gate-centric checks approve the transition, unaware that the control flow leading to it has been hijacked. 
Addressing this threat requires moving beyond gate verification to enforcing Control-Flow Integrity (CFI) on the path to the switch, as implemented in \sysname.

\section{System Overview} \label{sec:overview}

Existing intra-process isolation mechanisms face an isolation trilemma, forcing a trade-off between Scalability, Efficiency, and Security. Software-based approaches scale well but suffer from high overhead or weak security, while hardware-assisted ones (like MPK) are efficient but capacity-limited. \sysname addresses this by synthesizing complementary hardware capabilities into a hybrid architecture.

\noindent \textbf{Breaking the Scalability Limit.} To overcome the capacity constraints of single hardware primitives (e.g., POE's 7 domains), \sysname adopts a hierarchical composition strategy. It combines user-space POE, kernel-level PIE, and firmware-level GPC3 to instantiate domains across three tiers. By leveraging GPT's ability to map multiple physical address spaces (PAS), we effectively decouple the logical domain count from hardware limits, achieving unlimited scalability.

\noindent \textbf{Preserving Efficiency via Locality.} Hierarchical isolation introduces latency heterogeneity: switching deep tiers (PAS) is significantly costlier than user-level tiers (POE). We mitigate this by exploiting request locality. \sysname's scheduler employs an affinity-based allocation algorithm that groups sequential requests from the same connection into contiguous domains mapped to the same PAS. This maximizes execution residency in the fast L1-Zone (POE), masking the latency of rare privileged transitions.

\noindent \textbf{Enforcing Security under Adversarial Kernels.} To counter domain-switching abuse without relying on a trusted OS, \sysname anchors its root of trust in the minimal root-world Monitor. We enforce Code-Pointer Integrity (CPI) to secure the entire control-flow path leading to the switching trampoline, preventing borrowed-gate attacks. Furthermore, the Monitor offloads critical configuration tasks (e.g., PIE/POE setup) from the OS, ensuring isolation integrity even against a compromised kernel.

\begin{figure}[!t]
  \centering
  \includegraphics[width=\linewidth]{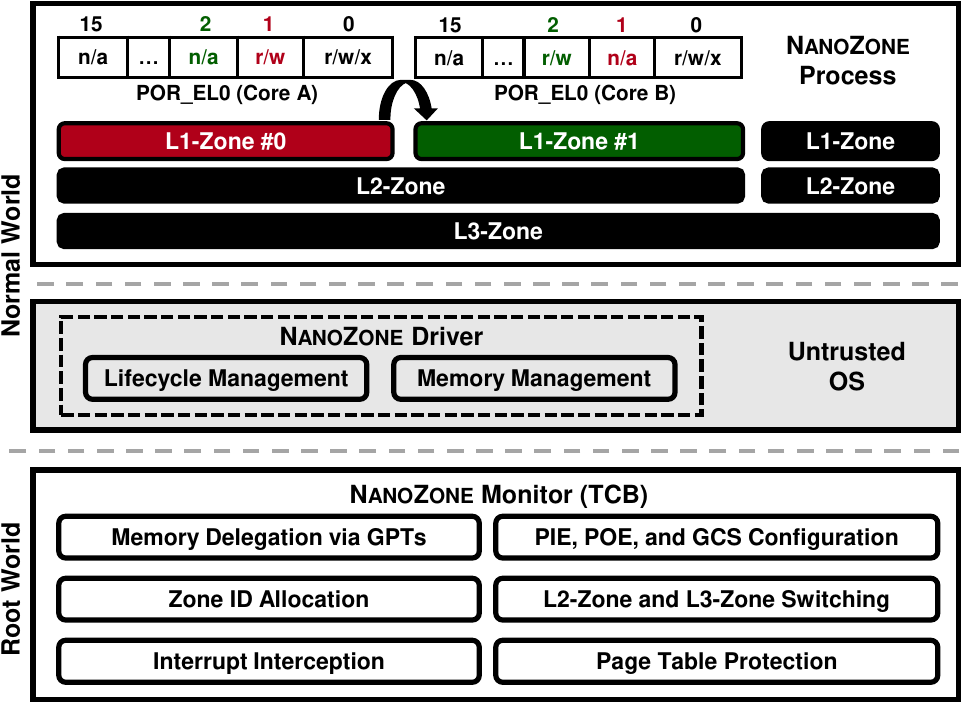}
  \caption{The architecture of \sysname.}
  \label{fig:architecture}
\end{figure}

\noindent \textbf{System Architecture.}
The architecture of \sysname is illustrated in \autoref{fig:architecture}. 
In our design, Arm CCA serves as the hardware substrate rather than the software deployment environment: \sysname repurposes CCA's GPT, GPC, and GPC3 primitives for isolation, but protected processes run in the normal world and never enter the realm world or use CVM-based isolation. \sysname divides the address spaces of protected processes into developer-defined domains across three tiers: POE domains (L1-Zone), PIE domains (L2-Zone), and GPT PASs (L3-Zone). The security of these domains is enforced by the Monitor in the root world. To further minimize the trusted software stack, \sysname adopts a delegation-based design that separates \emph{control} from \emph{ownership}. Specifically, we introduce a driver to the OS that manages the lifecycle of protected processes and memory allocation. While the OS can create, schedule, and destroy protected processes, it cannot access their data, code, page tables, or register states once the processes are delegated.

During compilation, we modify the LLVM compiler to identify all code pointers (e.g., function pointers, return addresses) for instrumentation protection, combined with post-compilation binary scanning to eliminate unsafe pre-inlined domain-switch or GCS store instructions (\cref{subsec:cfi}).

In the setup phase, the OS first loads the instrumented and scanned binaries and then invokes the driver to pre-allocate and map memory for isolation domains, along with their security control structures (e.g., page table and control-flow backups). The driver requests the Monitor to audit the loading procedure and delegates the pre-allocated physical memory to the protected process by configuring the OS's GPT, ensuring that the OS cannot access these memory regions. The Monitor then configures the protected process's GPT to isolate all L3-Zones from intra-process adversaries (\cref{subsec:reduction}). Next, the Monitor partitions the L3-Zone into L2-Zones and L1-Zones through PTE configurations (i.e., PIE and POE indexes) and marks the control‑flow‑backup pages with the GCS attribute (\cref{subsec:extend}). Finally, the Monitor initializes all registers associated with these features.

During runtime, the Monitor assigns isolation domain IDs for the current execution context. The user space facilitates L1-Zone context switching alongside CPI enforcement, which mitigates borrowed-gate attacks, without requiring privileged transitions to the Monitor. However, switching between L2-Zones and L3-Zones requires the Monitor to update the GPC3 and PIE registers (\cref{subsec:extend}). Throughout process execution, the Monitor intercepts each interrupt to enforce memory isolation and protect the register state. Before resuming the process, the Monitor verifies and synchronizes page table updates to prevent unauthorized modifications to PIE, POE, and GCS configurations (\cref{subsec:reduction}).

\noindent \textbf{Threat Model and Assumptions.}
\sysname's TCB is confined to the native root-world firmware and the security modules it adds. This firmware is loaded through CCA Secure Boot, so its integrity can be verified with the vendor's signature. We trust the underlying hardware to implement the required features (e.g., RME, PIE/POE, and GCS) correctly and assume a clean toolchain, meaning the application is built on a secure host. The normal-world OS (including our driver) and the software stacks in the realm and secure worlds are treated as untrusted.

\sysname protects against both privileged and intra-process adversaries. Privileged attackers may control the OS to remap domain memory, alter domain configurations, tamper with permission registers, inject malicious interrupts, forge domain‑switch requests, or exploit DMA devices. \sysname also accounts for Iago~\cite{checkoway2013iago} attacks that return malicious memory mappings. In parallel, intra-process adversaries may exploit memory bugs~\cite{durumeric2014matter} within the untrusted code to breach domain boundaries or hijack control flow~\cite{bletsch2011jump,shacham2007geometry}, enabling domain‑switching abuse (e.g., mid-trampoline injection, unauthorized entry, and borrowed-gate attacks). Finally, \sysname enforces mutual isolation so that a compromised domain cannot target other domains or escalate privilege to higher‑level software.

\sysname excludes scenarios where isolation domains intentionally leak sensitive data or contain internal vulnerabilities. Physical attacks~\cite{lee2020off,yitbarek2017cold,kim2014flipping}, microarchitectural side channels~\cite{lipp2018meltdown,kocher2020spectre}, denial-of-service (DoS) attacks, and Iago attacks beyond memory mapping are outside our threat model.

\section{\sysname Design} \label{sec:design}

\subsection{Isolation Domain Extension} \label{subsec:extend}

\sysname employs a standard code-centric workflow~\cite{liu2015thwarting,dinh2023capacity}: developers mark sensitive code, allocate domain memory, and insert call gates to thwart memory-safety exploits~\cite{durumeric2014matter}. 
Targeting server workloads, it also supports a data-centric model~\cite{lefeuvre2024sok} that isolates per-request contexts on demand. 
Fundamentally, \sysname redesigns the core mechanisms of memory configuration and permission switching, enabling unlimited domain scalability while strategically allocating IDs to minimize expensive privileged transitions.

\noindent \textbf{Extended Domain Configuration.}
With POE alone, the 3‑bit POE index (value 0 being reserved) limits the system to seven addressable domains. \sysname breaks past this cap in two ways: (1) it taps the remaining unused index encodings, and (2) it reuses the same encoding concurrently for separate, non‑overlapping physical memory regions.

\begin{figure}[!t]
  \centering
  \includegraphics[width=\linewidth]{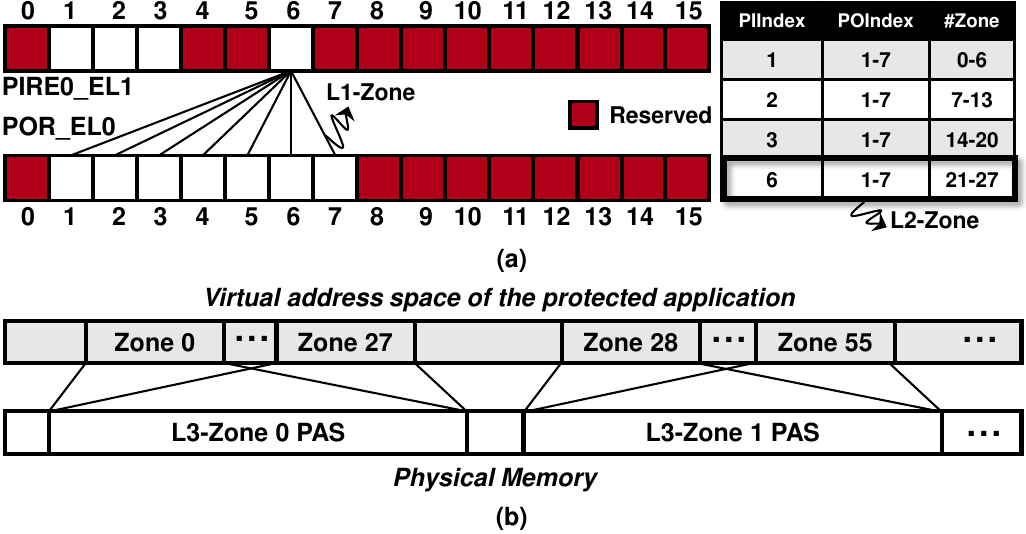}
  \caption{Structure of the domain extension: (a) shows the extension via PIE, while (b) depicts the extension via multi-PAS.}
  \label{fig:allocation}
\end{figure}

PTEs carry a 4‑bit PIE field for base permissions in addition to the 3‑bit POE field for overlay permissions. While most PIE index encodings are already in use (six for common permissions, two for GCS, and four for kernel), we observe that there remain four unused encodings. As shown in \autoref{fig:allocation}(a), \sysname combines PIE and POE indexes to form the domain ID, expanding the addressable domains to 28. Isolation between these expanded domains is enforced in two cases. 
(1) Domains within the same L2-Zone. For example, Domain-A (\texttt{PI:6, PO:1}) and Domain-B (\texttt{PI:6, PO:2}) share the same base permission (index 6) in the PIE permission register (\texttt{PIRE0\_EL1}). However, their overlay permissions differ in the POE permission register (\texttt{POR\_EL0}). When executing in Domain-A, \texttt{overlay perm1} is enabled and \texttt{overlay perm2} is forced to 0 (overriding \texttt{base perm6}), which blocks any access to Domain-B. (2) Domains across different L2-Zones with a shared POE index. Consider Domain-A (\texttt{PI:6, PO:1}) and Domain-C (\texttt{PI:3, PO:1}). Both domains enable \texttt{overlay perm1}, but Domain-C's base permission (index 3) remains disabled. As a result, even though they share the same POE index, no lateral access is allowed from Domain-A into Domain-C. Notably, PIE indexes with fixed semantics (for example, the default index 4) are never reused, since their base permissions cannot be masked.

The 28 POE-PIE domains still fall short of the scalability needed for multi-tenant servers; therefore, \sysname turns to another CCA feature, the GPT, which partitions physical memory into multiple PASs beyond the page-table layer. A simple idea is to map each POE‑PIE domain to its own PAS, so a domain is identified by ⟨PAS, PIE, POE⟩; see \autoref{fig:allocation}(b). This raises a new problem: the requesting core must gain exclusive access to its target PAS while all other cores are blocked. Existing schemes give every PAS a private access‑control table (e.g., EPK~\cite{gu2022epk} maintains extra EPTs), but this per-PAS-per-table model inflates memory and incurs creation latency (both detailed in \cref{subsec:microbenchmarks}), while being constrained by hardware limits such as 512 EPTs per VM. 
\sysname keeps all PASs inside a single process's GPT and uses the GPC3 bypass‑window feature. During setup each PAS is marked \texttt{no‑access}. When a core needs to access a particular PAS, the Monitor programs that core's \texttt{GPCBW\_EL3} register to disable the GPC only for the required range; other cores still fail the check. Since the Monitor pre‑defines every legal window, attackers cannot open arbitrary ones.

Keeping a single GPT also strengthens isolation under an untrusted OS. Per-PAS-per-GPT schemes grant a core access by rewriting a region's attribute to an accessible state. The correctness of such a scheme then depends on validating, at every rewrite, that the region is not cross-world memory; a single omitted check exposes another world's memory to a protected application. We observed a concrete instance of this risk in the implementation of Shelter~\cite{zhang2023shelter}, which we have reported to its authors. In \sysname, by contrast, a protected PAS is never rewritten and remains \texttt{no-access} throughout; access is determined solely by the Monitor-exclusive \texttt{GPCBW\_EL3} register.

\begin{figure}[!t]
  \centering
  \includegraphics[width=\linewidth]{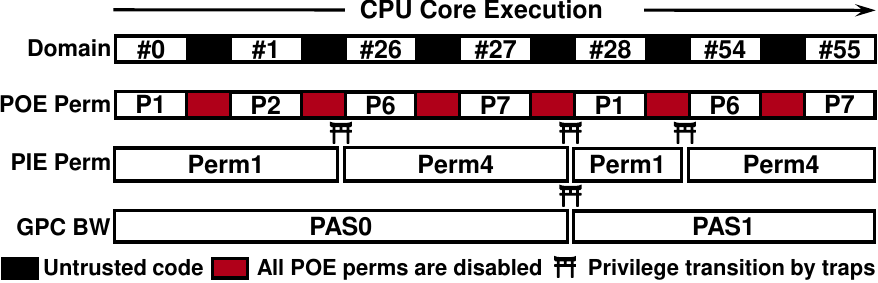}
  \caption{Intra-process permission switching in the \sysname lifecycle.}
  \label{fig:lifecycle}
\end{figure}

\noindent \textbf{Domain Switching.}
Like prior works~\cite{schrammel2020donky,gu2022epk,yuan2024lightzone,gu2022hardware}, \sysname packages its domain-switching code into a dedicated trampoline that orchestrates transitions across the three isolation levels. 
Permission grants are usually trap-free; each involves enabling the corresponding overlay permission by writing \texttt{POR\_EL0}, but only transitions that cross L2- or L3-Zone boundaries require a privileged trap. To revoke access, user code simply clears all overlay permissions; even if the PIE base permission or bypass window stays active, the cleared overlay ensures untrusted code cannot reach isolation domains (see \autoref{fig:lifecycle}). For an L2 switch, the Monitor writes \texttt{PIRE0\_EL1} to disable the old base permission and enable the new one. For domains within an L2-Zone that require special privileges, we set their base permission to full \texttt{R/W/X} and use POE overlays for fine-grained access control, avoiding unnecessary traps. Switching between L3-Zones updates \texttt{GPCBW\_EL3} to point the bypass window at the new PAS.

\noindent \textbf{Domain ID Allocation Optimization.}
L2 and L3 switches trigger privileged traps, so \sysname lowers their frequency by redesigning domain allocation. First, it follows a per-connection-per-domain policy, keeping consecutive requests from the same connection in the same domain and limiting them to inexpensive L1 switches. Second, servers like Memcached hand connections to worker threads in round‑robin fashion; if domains were allocated in that same order, each thread would own a scattered set spanning many L2 and L3 zones, sharply raising the trap rate. \sysname instead grants each thread a block of 28 consecutive domains that all map to the same PAS. Threads still process connections sequentially, but each one lands in a locally contiguous domain. Thus only a small share of requests must perform L2 and L3 switches (see \cref{subsec:microbenchmarks} for hit-rate details).

\subsection{Code-Pointer Integrity} \label{subsec:cfi}

The root cause of domain-switching abuse lies in two vectors. (1) pre-inlined switching instructions embedded in the binary, and (2) control-flow hijacking attacks that exploit existing switching logic, whether via returns/jumps to the trampoline or direct mid-trampoline execution of switching instructions. Notably, merely validating that an indirect branch reaches a legitimate landing pad is insufficient in our threat model. An adversary may still corrupt control-flow data in untrusted code and redirect execution to an otherwise legitimate switch site, i.e., a borrowed-gate attack. To counter these attacks, \sysname first performs binary scanning to eliminate pre-inlined switching instructions (e.g., unauthorized \texttt{POR\_EL0} updates), ensuring that only the mapped trampoline can change permissions. Second, \sysname blocks switching reuse by enforcing CPI on return addresses and function pointers, thereby protecting the entire control-flow path leading to a domain switch.

\noindent \textbf{Return Address Protection.}
\sysname employs the off-the-shelf GCS feature to secure return addresses. On each branch-with-link, the CPU pushes the return address to the protected GCS region and, on return, checks it against the value popped from that stack. The enabling and configuration of GCS, such as setting the stack pointer to a securely pre-delegated memory area, are managed by the Monitor.

\begin{figure}[!t]
  \centering
  \includegraphics[width=\linewidth]{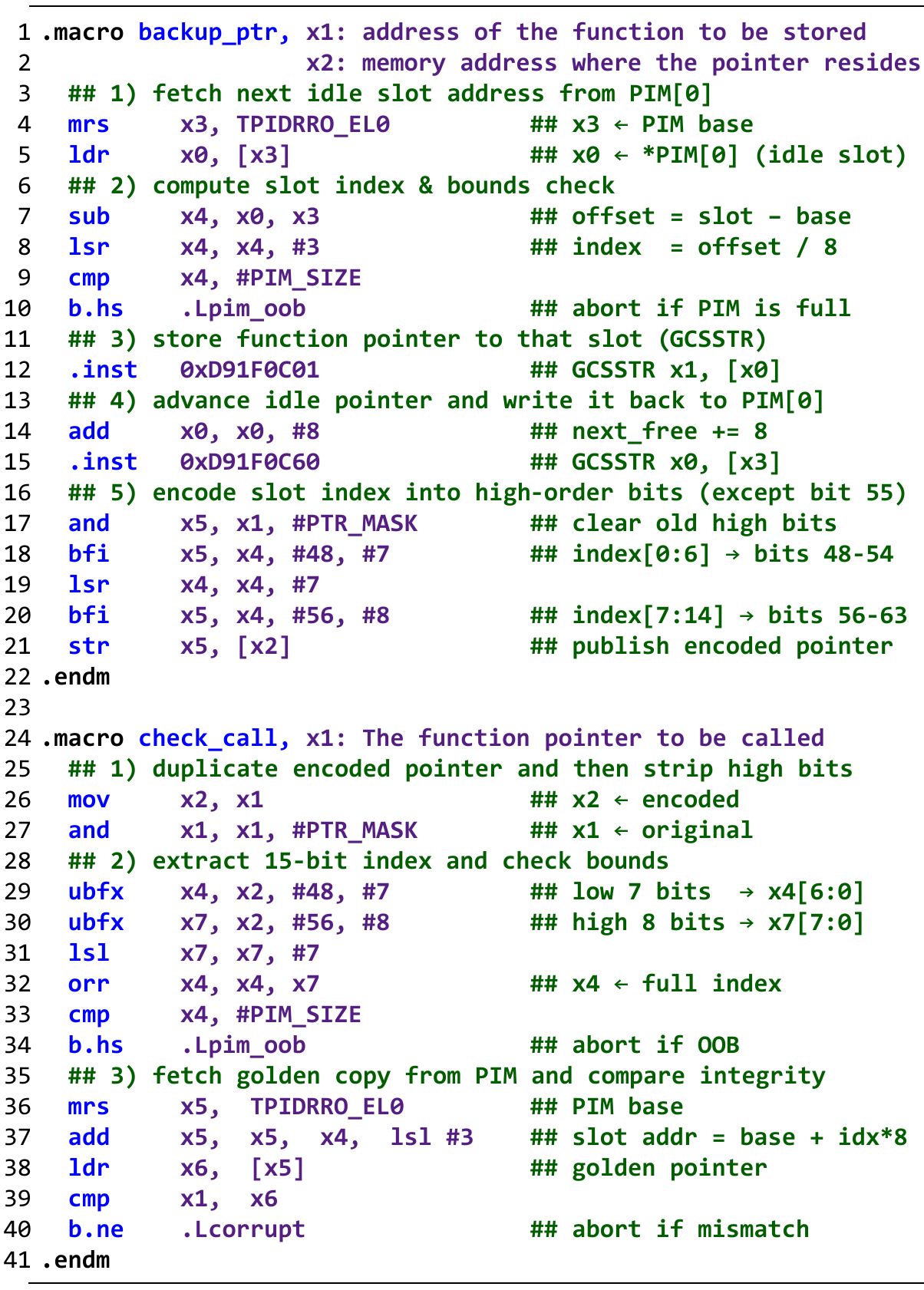}
  \caption{Instrumentation code for pointer backup and call verification (type-ID handling omitted).}
  \label{fig:pim}
\end{figure}

\noindent \textbf{Function Pointer Protection.}
The core idea behind our function pointer protection is to create a set of legitimate landing pads (indirect jump targets), back them up into the PIM region (a \sysname-managed read-only memory area with GCS attributes) and enforce pre-jump integrity checks. Despite conceptual simplicity, it faces three critical challenges: (1) ensuring threads securely acquire their PIM base address, (2) optimizing the efficiency of backup storage and lookup during checks, and (3) preventing reuse attacks that exploit previously backed-up legitimate targets.

We observe that \texttt{TPIDRRO\_EL0} is a \emph{read-only}, user-accessible per-thread register. The OS commonly uses this zero-latency path to expose values that must remain constant for a thread's lifetime. Following the same practice, \sysname stores the PIM base address in \texttt{TPIDRRO\_EL0}, shielding it from intra-process tampering.

For efficient backups (\autoref{fig:pim}, \texttt{backup\_ptr}), the first PIM entry holds the address of the next free slot, so user-mode code stores the backup with one non-privileged \texttt{gcsstr} and advances the slot address with another. Note that during binary scanning we also strip out any pre-inlined \texttt{gcsstr} so that only our instrumentation can invoke it. This does not affect the GCS protection of ordinary applications. The return-address shadow stack is maintained by hardware on \texttt{BL} and \texttt{RET} with no explicit store, so a normally compiled application contains no \texttt{gcsstr}; the instruction performs an explicit GCS write, a capability disabled by default and used only by low-level components such as checkpoint/restore. Stripping it therefore removes an abuse path through which untrusted code could forge PIM entries, without affecting any application's return-address integrity. Lookup is also constant-time (\autoref{fig:pim}, \texttt{check\_call}): \sysname embeds the slot index into the unused high-order bits of the pointer during backup. This index travels with the pointer as it is copied, avoiding extra memory or propagation tracking. To eliminate redundant storage during pointer propagation, \sysname checks the high-order bits: if they already contain a valid index, the pointer has been backed up previously and the operation is skipped. \sysname retrieves the backup directly from the PIM using the base address and index for comparison with the target. Before utilizing the index, range checks ensure accesses stay within the PIM boundaries, preventing attackers from retrieving forged backups via out-of-bounds indexes (see \cref{subsec:microbenchmarks} for detailed CPI costs).

Attackers could try to reuse legitimate landing pads. To mitigate this, we introduce pointer-type tags for function pointers. During compilation, we extract each function signature at the LLVM IR level, generate a truncated 64-bit hash (serving as a unique type ID), and embed this ID directly into IR instructions to preserve type context across the compilation pipeline. When backing up function pointers, the ID is stored as metadata alongside the pointer in the PIM. At call sites, both the pointer and its associated ID are retrieved, and the ID is validated against the expected type hash, thereby hindering gadget chaining.

\sysname also needs to handle some corner cases. Global function pointers are initialized at compile time, bypassing the runtime backup. We modify the linker script to place all initialized global function pointers into a dedicated segment. During binary loading, this segment is copied into the PIM, and each pointer is stamped with its corresponding PIM index in the high-order bits. Additionally, to address backup omissions from assignments to universal pointers (e.g., \texttt{void*}), we insert extra checks to detect whether the source address is a function pointer that has been typecast (i.e., via \texttt{bitcast}). If so, the pointer is backed up, and the type ID is derived from the original source type. LLVM IR's strong typing guarantees that any later call through the universal pointer must cast it back to a concrete function-pointer type, and that cast exposes the type we need. Further propagations (even nested ones) need no extra handling because the pointer has already been recorded.

\begin{figure}[!t]
  \centering
  \includegraphics[width=\linewidth]{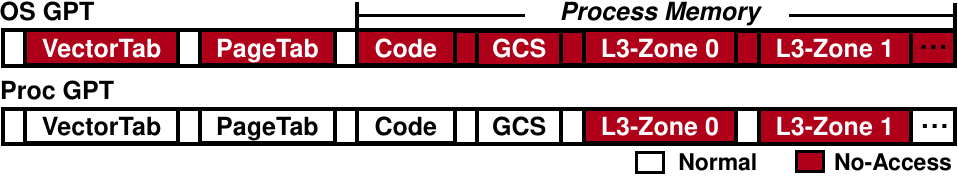}
  \caption{A dual-GPT design blocking OS access to secure memory.}
  \label{fig:multi-gpt}
\end{figure}

\subsection{OS Privilege Reduction} \label{subsec:reduction}

To prevent the OS from breaking the established isolation, \sysname restricts its privileges in three ways: (1) it denies the kernel access to secure memory regions, (2) it blocks any kernel changes to sensitive interrupt context, and (3) it forbids unauthorized updates to page-table mappings or permission configurations of POE, PIE, and GCS pages.

\noindent \textbf{Memory Isolation.}
Managing access at the virtual memory layer, such as hooking page tables for the kernel and all other processes to block alias attacks, is both complex and heavyweight. Prior works~\cite{zhang2023shelter,zhourcontainer} shift control to the physical memory layer with GPTs, delivering a simpler and more efficient solution. \sysname therefore maintains two distinct GPTs: an OS GPT and a proc GPT (see \autoref{fig:multi-gpt}). The OS GPT denies the kernel any access to process memory, including domain pages, the GCS and PIM regions, and executable code, and also revokes access to interrupt-vector tables and page tables. Once the driver allocates these regions, the Monitor marks them \texttt{no-access} before handing them to the process. The proc GPT, by contrast, leaves everything accessible except the PASs that make up L3-Zones, which stay blocked until a bypass-window switch occurs (\cref{subsec:extend}). Other sensitive areas (e.g., read-exec code) remain protected by ordinary page-table permissions. To ensure both GPTs coexist safely, the Monitor checks that delegated memory never overlaps existing regions and, when the realm world or secure world requests pages, updates the proc GPT to prevent isolation domains from reaching them. For system calls that pass arguments through a process buffer, the Monitor allocates and maintains shared memory to keep the OS from touching process memory directly.

\noindent \textbf{Interrupt Interception.}
To isolate process state from the OS, \sysname intercepts all user-to-kernel transitions (see \cref{subsec:microbenchmarks} for detailed costs). A pre-delegated interrupt vector (see \autoref{fig:multi-gpt}) ensures that control first enters EL3 before reaching the kernel. The Monitor swaps the proc GPT for the OS GPT, saves and sanitizes sensitive context (POE/PIE permission registers and GCS/PIM pointer registers), and, just before handing control back, restores this context while re-enabling PIE, POE, and GCS so the OS cannot silently disable them. However, we observe that domain switches occur far more often than interrupts, so forwarding each L2- or L3-Zone transition through a two-stage trap (first to the vector table, then to the Monitor) would be too expensive. Capturing PIE register accesses from EL3 (i.e., via \texttt{SCR\_EL3}) is also impractical due to overly coarse control; once enabled, all POE register accesses would likewise trap to the Monitor, undermining our \emph{\textbf{efficiency goal}} (\cref{sec:overview}). Instead, \sysname exploits the Armv8.5-A RNG TRAP feature: any user-space read of \texttt{RNDR} or \texttt{RNDRRS} traps directly to the Monitor. \sysname co-opts these registers as lightweight triggers for L2 and L3 switches, entirely bypassing kernel mode. Since RNG trapping also applies to the OS, the Monitor validates each request by checking the current GPT base, blocking domain-switching abuse.

\noindent \textbf{Page Table Protection.}
As noted earlier, \sysname configures the OS GPT to deny the OS access to process page tables (see \autoref{fig:multi-gpt}). When a page fault occurs, the OS forwards the mapping request to the Monitor, which performs the actual page table updates. The Monitor enforces strict checks to ensure that: (1) newly mapped pages are non-executable, and (2) existing secure memory mappings and their configurations (i.e., PIE/POE indexes in PTEs) are not modified. Rule (1) is deliberately absolute. All code pages, including shared libraries resolved by the dynamic linker, are mapped and audited during the setup phase. After delegation, the Monitor denies any transition of a page to executable. Runtime code loading such as \texttt{dlopen} is therefore unsupported by design. This closes the runtime code-injection surface entirely. We note this restriction in \cref{sec:discussion}. Additionally, \sysname defends against memory-mapping Iago attacks~\cite{checkoway2013iago}, such as malicious \texttt{mmap}, by verifying that new mappings do not overlap with existing ones.

\section{Implementation} \label{sec:prototype}

At the time of writing, no commercial SoC, including the Cortex-X925~\cite{2024armx925}, supports the Arm extensions our work depends on. Therefore, we implemented: (1) a functional prototype on the Arm Fixed Virtual Platform (FVP)~\cite{arm2023fvp}, which natively supports these features, to evaluate security properties; and (2) a performance prototype on physical Arm SoCs, where the missing features are emulated, to assess execution overhead.

\noindent \textbf{Functional Prototype.} 
\sysname instruments source code with LLVM 16.0.0 to secure function pointers. To locate where a function pointer is stored, it iterates over every \texttt{StoreInst} and \texttt{MemCpyInst}, since the latter often initializes struct fields that are function pointers. With \texttt{getPointerElementType}, it checks the pointee type: if the destination is a function pointer, or the source was a function pointer before a \texttt{BitCastInst}, it inserts backup-protection code. \sysname then analyzes each \texttt{CallInst} to identify call sites. When the callee is a runtime \texttt{Value} (e.g., from a \texttt{LoadInst} or \texttt{BitCastInst}), it inserts code that verifies the pointer against its backup. Both backup and verification derive a function-type ID by hashing the type with SHAKE-128, mitigating reuse attacks.
While requiring recompilation, this toolchain is essential for intra-process security; it automates CFI insertion and strips unsafe instructions, eliminating manual refactoring. 
This workflow integrates seamlessly into ``Secure DevOps'' pipelines, enabling tenants to transparently harden workloads (e.g., Nginx) during the build process.

Since GPC3 supports only contiguous bypass windows, the \sysname kernel driver leverages the Contiguous Memory Allocator (CMA) to allocate a contiguous block of PAS for the L3-Zone and remaps it into the process's address space. To support application development, \sysname exposes a minimal API including \texttt{zone\_alloc}/\texttt{zone\_free}, \texttt{zone\_begin}/\texttt{zone\_end}, and memory management via \texttt{zone\_mmap}/\texttt{zone\_munmap}.
Existing domains can request the Monitor (Trusted Firmware-A 2.9.0) to allocate the pre-mapped memory using these interfaces.
Notably, to ensure high performance without burdening the developer, \texttt{zone\_alloc} implements the affinity-based allocation strategy: it automatically assigns a domain ID that maps to the requesting thread's local PAS, thereby defaulting subsequent accesses to efficient L1-Zone switches. Each interface call is wrapped in an \texttt{svc} instruction, which traps into a protected vector table that immediately issues an \texttt{smc} to enter the Monitor. Regular interrupt traps follow a similar path to reach the Monitor. Before the exception returns to the process, the Monitor re-enables PIE and POE in \texttt{TCR2\_EL1}, GCS in \texttt{GCSCRE0\_EL1}, and RNDR trapping in \texttt{SCR\_EL3}, ensuring these features stay active in user space.
During our development, the latest available Linux kernel (v6.7.0) officially supported only the PIE. Support for POE and GCS had not yet been upstreamed, so we applied patches for these features from Patchwork~\cite{2024linuxpatchwork}, which were still under review at the time. It is worth noting that starting from v6.13-rc1, these features have been merged into the mainline. To enable these features in the FVP, the required boot parameters must be enabled: 
\texttt{has\_rndr\_trap},
\texttt{has\_permission\_overlay\_s1},
\texttt{rme\_support\_level}, 
\texttt{has\_rme\_gpc3},
\texttt{has\_gcs}, 
and \texttt{has\_permission\_indirection\_s1}.

\noindent \textbf{Performance Prototype.} We built the performance prototype on a Rockchip RK3399 (Armv8) board~\cite{2024radxarockpi}. Its implementation mirrors the functional prototype, except that the required hardware features are emulated as follows:
For L1-Zone switches we emulate accesses to \texttt{POR\_EL0} with \texttt{TPIDR\_EL0}. Since both are unprivileged system registers, the emulation latency closely mirrors real hardware behavior. L2-Zone switches repurpose \texttt{AFSR0\_EL1} as \texttt{PIRE0\_EL1}, while L3-Zone bypass windows substitute \texttt{ACTLR\_EL3} for \texttt{GPCBW\_EL3}. Since the hardware RNG-trap is absent, L2 and L3 transitions are also driven by \texttt{svc} and \texttt{smc}; actual hardware overhead will therefore be lower than our emulation. We also map \texttt{AFSR0\_EL3} and \texttt{AFSR1\_EL3} to \texttt{GPCCR\_EL3} (GPC control) and \texttt{GPTBR\_EL3} (GPT base) to measure GPC-setup and GPT-switching costs. PIE domains are emulated by reusing available PTE bits in the direct-memory model, while POE domains use reserved PTE fields. The \texttt{gcsstr} instruction is replaced by a standard \texttt{str}, and \texttt{tlbi paallos}, which normally flushes only cached GPT entries, is conservatively replaced by a full TLB invalidation. 
Since GPTs are memory-mapped, their creation, maintenance, and teardown overhead is modeled identically in both the functional and performance prototypes.

\section{Security Evaluation} \label{sec:security_evaluation}

\subsection{TCB Analysis} \label{subsec:tcb}

\sysname's TCB is confined exclusively to the Monitor module residing in the root world (EL3), totaling approximately 3K LoCs. The kernel driver ($\sim$1K LoCs) operates outside the trust boundary and is treated as untrusted, as its compromise does not violate the core isolation guarantees enforced by the Monitor. The compile-time toolchain ($\sim$500 LoCs) is trusted but limited to the build phase, as discussed in the threat model (\cref{sec:overview}).

\noindent \textbf{TCB Scope and Functional Breakdown.}
The Monitor module's 3K LoCs encompass six well-scoped responsibilities: (1) GPT memory delegation and lifecycle management, (2) PIE/POE/GCS register configuration, (3) domain ID allocation, (4) L2/L3-Zone switching, (5) interrupt interception and context sanitization, and (6) page table protection. Each component has a narrow, verifiable interface, making the TCB small enough to be a realistic target for manual auditing or formal verification. Crucially, by anchoring all security logic in EL3, \sysname eliminates the need to trust the OS, the RMM, or any realm-world software, which is a critical property given that all of these are treated as adversarial in our threat model (\cref{sec:overview}).

\subsection{Theoretical Security Analysis} \label{subsec:analysis}

We analyze security by listing attack vectors from the threat model and detailing how \sysname mitigates them.

\noindent \textbf{Untrusted OS}.
At startup, the OS might tamper with a program's image or map several binaries onto overlapping pages. The Monitor counters this by validating the load address, rejecting overlapping mappings, and auditing code pages before locking them against further modification. Additionally, verifying the image content against the tenant's intended binary requires a trusted reference measurement, whose integration we discuss in \cref{sec:discussion}. To prevent races between concurrent delegation requests, all GPT operations are synchronized with a spinlock.

At runtime, the OS GPT blocks the kernel from touching a protected process's page tables; every mapping request is vetted by the Monitor. The kernel therefore cannot place an isolation-domain or GCS/PIM page outside the assigned L3-Zones or change a page's L1- or L2-Zone assignment by editing its PTE indexes. From EL1, the kernel might still try to (1) tweak POE/PIE permission registers to force an L1/L2 switch, (2) redirect the GCS stack pointer, (3) forge an interrupt-return address, or (4) disable the isolation features. All attempts fail because control always returns to EL3 first, where the Monitor restores the registers and re-enables the features (\cref{subsec:reduction}). Any effort to trigger an L2 or L3 switch by touching the RNG TRAP register under the OS GPT is likewise rejected; the Monitor allows access only when execution is under the proc GPT. Iago attacks~\cite{checkoway2013iago} that manipulate the return values of memory-mapping calls such as \texttt{mmap} are thwarted because the Monitor refuses overlapping mappings, and GCS/PIM make a follow-on control-flow hijack even harder. Running in the highest-privilege root world, the Monitor is beyond the kernel's reach: the OS cannot bypass its interrupt interception, GPT memory control, or change EL3 settings (e.g., GPT base or L3-Zone bypass windows). Finally, since GPT entries and bypass windows can reside in the TLB, the Monitor flushes the TLB on each interrupt or L3 switch and configures the \texttt{TTBR} \texttt{CnP} bit to prevent cross-core sharing.

Peripheral-side attacks deserve equal caution. Following prior CCA systems~\cite{zhang2023shelter,wang2024cage,sang2025portal,huang2026hivetee}, we configure the SMMU's GPT to enforce GPC for devices, blocking DMA access to protected memory. Additionally, we flush its translation caches, mirroring CPU-side hardening, to prevent TLB leaks through stale entries or bypass windows.

\noindent \textbf{Intra-process Adversaries}.
Memory isolation for each domain is enforced via a three-tier regime (POE/PIE/GPC3), thwarting even arbitrary out-of-bounds reads/writes by untrusted code (\cref{subsec:extend}). Backed-up return addresses and function pointers stop control-flow hijacks such as ROP~\cite{shacham2007geometry} or JOP~\cite{bletsch2011jump}, blocking any attempt to abuse domain switches; reuse attacks invoking legitimate targets are limited to same-type targets by type ID validation at call sites (\cref{subsec:cfi}).

\noindent \textbf{Malicious Domains}.
A domain runs at EL0 in the normal world, so although it can access its own private memory, it lacks the privilege to access the other worlds or the normal-world kernel. Note that the OS page table can still prevent lateral escalation into other user processes. To address potential collusion with the OS, the Monitor checks that each OS-supplied L3-Zone region does not overlap existing memory and synchronously updates any memory delegations requested by other worlds in the proc GPT (\cref{subsec:reduction}).

\subsection{CVE Mitigation Analysis} \label{subsec:cve}

Beyond theoretical analysis, we evaluated \sysname's practical defenses against known CVEs. On Arm FVP, \sysname blocks all vulnerabilities in \autoref{tab:cve}, including five memory-safety flaws and five control-flow hijacks. 

\begin{table}[t]
  \centering
  \caption{CVEs used for mitigation analysis.}
  \label{tab:cve}
  \small
  \begin{adjustbox}{max width=\linewidth}
  \begin{tabular}{@{}ll@{}}
    \toprule
    \textbf{CVE-*} & \textbf{Description} \\
    \midrule

    \multicolumn{2}{@{}l}{\textit{Memory-Safety Vulnerabilities}}\\[2pt]
    2014-0160  & Out-of-bounds read in OpenSSL's TLS heartbeat. \\
    2017-2800  & Off-by-one overwrite in wolfSSL's X.509 parser. \\
    2017-18922 & Heap overflow in LibVNCServer's WebSocket decoder. \\
    2022-24834 & Heap overflow in Redis's cjson library. \\
    2023-3138  & Out-of-bounds write in libX11's \texttt{InitExt}. \\

    \addlinespace[3pt]
    \multicolumn{2}{@{}l}{\textit{Control-Flow Hijacking}}\\[2pt]
    2013-2028  & Return-address overwrite in Nginx's chunked parser. \\
    2015-7805  & Function-pointer overwrite in libsndfile's AIFF parser. \\
    2016-5314  & Function-pointer overwrite in LibTIFF's PixarLogDecode. \\
    2021-44486 & Function-pointer overwrite in YottaDB's \texttt{op\_write}. \\
    2024-22857 & Function-pointer overwrite in zlog's \texttt{zlog\_rule\_new}. \\
    \bottomrule
  \end{tabular}
  \end{adjustbox}
\end{table}

Memory bugs typically involve out-of-bounds reads (e.g., CVE-2014-0160) or writes (e.g., CVE-2023-3138) targeting sensitive data. 
\sysname places secret pages in dedicated isolation domains, ensuring any unauthorized access triggers a POE/PIE fault. 
For instance, in Heartbleed (CVE-2014-0160), OpenSSL's \texttt{tls1\_process\_heartbeat} fails to validate payload length, allowing a 64 KB over-read. 
Under \sysname, since private keys reside in a protected domain, this over-read provokes a hardware fault rather than leaking secrets.

Control-flow hijacks exploit corruption to overwrite return addresses (e.g., CVE-2013-2028) or function pointers (e.g., CVE-2021-44486). 
Consider CVE-2024-22857: a heap overflow in zlog's \texttt{zlog\_rule\_new} allows a crafted configuration to overflow the \texttt{record\_name} buffer and clobber the adjacent \texttt{record\_func} pointer. 
\sysname prevents this by validating function pointers against PIM-stored backups on every indirect call, immediately terminating execution if tampering is detected.

We designed two attacks to evaluate \sysname's resilience to privileged adversaries, and its defenses blocked both attempts.
(1) We mapped the process page tables into kernel space and altered the domain's PTEs, changing the POE and PIE indexes so that the pages would lie in a region with ordinary read/write permissions. The change immediately triggered a GPF.
(2) While handling \texttt{getpid}, we modified the POE and PIE permission registers in the kernel to grant read access to a domain that should have been private. When control returned to user mode, code outside the domain still could not access its memory because the Monitor had already restored the correct register values.

\section{Performance Evaluation} \label{sec:performance_evaluation}

\subsection{Experimental Setup} \label{subsec:setup}
We evaluate \sysname on a Rockchip RK3399 board (Armv8-A) with two Cortex-A72 cores (1.8 GHz) and four Cortex-A53 cores (1.4 GHz), a standard configuration in prior works~\cite{zhang2023shelter,zhourcontainer,wang2024cage}. 
To eliminate variability from the SoC's big.LITTLE design, we lock the Cortex-A72 cores at 1.8 GHz and disable the little cores.
Note that core pinning is employed strictly for benchmarking stability; \sysname naturally supports standard OS scheduling, as the Monitor transparently manages hierarchical isolation states during context switches.
Our evaluation covers two vectors:

\noindent \textbf{Micro-benchmarks.}
We first evaluate the efficiency of our bypass-window optimization against per-PAS baselines. 
Subsequently, we dissect the overheads of domain switching, CPI enforcement, and OS monitoring. 
Specifically, we measure the latency and hit rate of our three-tier switching logic, quantify CPI costs using SPEC CPU2017~\cite{2025speccpu2017}, and assess OS privilege de-escalation impact via lmbench~\cite{2025lmbench}.

\noindent \textbf{Real-world Applications.}
We validate \sysname's practicality by hardening three systems: (1) Nginx (isolating SSL keys), (2) Memcached (partitioning K-V pairs), and (3) a persistent NVM data store.

\subsection{Evaluation on Micro-benchmarks} \label{subsec:microbenchmarks}
 
\noindent \textbf{Initialization Optimization.}
To quantify the benefits of our bypass-window design during setup, we reimplemented the state-of-the-art per-PAS-per-GPT scheme from CAGE~\cite{wang2024cage} as a baseline. Both approaches construct the same OS GPT and proc GPT (\autoref{fig:multi-gpt}), but diverge thereafter. In the baseline, each new L3-Zone requires cloning the proc GPT, marking its PAS as normal (i.e., accessible), and assigning the copy to that zone. Note that the Level 1 (L1) tables in the proc GPT can be reused and each PAS is referenced by a single block descriptor in the Level 0 (L0) table, only the L0 table needs to be duplicated per zone. In contrast, \sysname always reuses the original two GPTs; each core simply programs its bypass-window register to open the PAS. Under a 25-zone workload, this saves 100 KB of DRAM compared to the baseline, while the performance overhead of L0-table duplication in the baseline is negligible (0.02\%).

\begin{table}[t]
  \caption{Cost of basic \sysname operations.}
  \label{tab:operation_cycle}
  \centering
  \small
  \begin{adjustbox}{max width=\linewidth}
  \begin{tabular}{@{}lccccccc@{}}
    \toprule
    \multirow{2}{*}{\textbf{Operation}} &
    \textbf{L1} &
    \textbf{L2} &
    \textbf{L3} &
    \textbf{Ptr.} &
    \textbf{Ptr.} &
    \multirow{2}{*}{\textbf{getpid}} &
    \textbf{Hooked} \\
    & \textbf{Switch} & \textbf{Switch} & \textbf{Switch} & \textbf{Backup} & \textbf{Check} & & \textbf{getpid} \\
    \midrule
    \textbf{Cycles} & 74.13 & 6,169.47 & 6,173.36 & 18.02 & 11.07 & 725.36 & 6,533.63 \\
    \bottomrule
  \end{tabular}
  \end{adjustbox}
\end{table}

\begin{table}[t]
  \caption{Average domain-switching overhead (cycles) and switching-level hit rates.}
  \label{tab:switch_cycle}
  \centering
  \small
  \begin{adjustbox}{max width=\linewidth}
  \begin{tabular}{@{}lcccccc@{}}
    \toprule
    \textbf{Domain Num.}\textsuperscript{1} & 7 & 14 & 28 & 112 & 168 & 224 \\
    \midrule
    Avg. Switch Cost           & 86.00   & 198.42 & 255.88 & 300.63 & 303.06 & 303.43 \\
    L1-Zone Hit Rate\textsuperscript{2} & 100.00\% & 98.21\% & 97.32\% & 96.65\% & 96.58\% & 96.52\% \\
    L2-Zone Hit Rate\textsuperscript{2} & 0.00\%   & 1.79\%  & 2.68\%  & 2.68\%  & 2.68\%  & 2.68\%  \\
    L3-Zone Hit Rate\textsuperscript{2} & 0.00\%   & 0.00\%  & 0.00\%  & 0.67\%  & 0.74\%  & 0.78\%  \\
    \bottomrule
  \end{tabular}
  \end{adjustbox}

\vspace{2pt}
\begin{minipage}{0.97\linewidth}
\footnotesize
\raggedright
\textsuperscript{1} Domain Number refers to the number of domains assigned per core.\\
\textsuperscript{2} Hit Rate refers to the proportion of switches whose highest switching level is L1, L2, or L3.
\end{minipage}
\end{table}

\noindent \textbf{Domain Switching Cost.}
To quantify base overheads, we used the PMU to measure CPU cycles over one million iterations (\autoref{tab:operation_cycle}). 
The results highlight a stark contrast: a lightweight L1 switch consumes only 74.13 cycles, whereas privileged L2 and L3 transitions exceed 6,100 cycles. This disparity motivates our locality-aware allocation strategy to maximize inexpensive L1 transitions.
We evaluated switching overhead using \texttt{twemperf}~\cite{twemperf2024} and varied the number of isolation domains while maintaining a constant request rate. Domain switches were strictly determined based on our allocation strategy.
With up to 7 domains, the system maintains a 100\% L1 hit rate, incurring just 86 cycles per switch (\autoref{tab:switch_cycle}). 
As we scale to 28 domains, L2 transitions become necessary but remain infrequent; the L1 hit rate stays high at 98.21\%, keeping the average cost to 256 cycles. 
Even at a massive scale of 224 domains, L3 switches occur only 0.78\% of the time.
Consequently, the average switching overhead stabilizes at roughly 300 cycles, which is less than half the cost of a standard \texttt{getpid} system call.
We capped the evaluation at 224 domains because the performance trend stabilizes beyond this point, and this scale already exceeds the empirical evaluation of prior solutions: LightZone~\cite{yuan2024lightzone} evaluated a maximum of 128 domains despite its higher theoretical limit.

\noindent \textbf{CPI Enforcement Overhead.}
\autoref{tab:operation_cycle} confirms the efficiency of our CPI primitives (\cref{subsec:cfi}): both backup and verification operations complete in under 20 cycles, benefiting from purely user-space memory accesses. 
To assess the impact on high-stress workloads, we utilized SPEC CPU2017~\cite{2025speccpu2017}. Due to the board's 4 GB RAM constraint, we restricted our evaluation to the C benchmarks from the SPEC rate2017 suite. 
We tested three configurations: (1) GCS protection for return addresses, (2) PIM backup for function pointers, and (3) integrated protection for both.

\begin{figure}[!t]
  \centering
  \includegraphics[width=\linewidth]{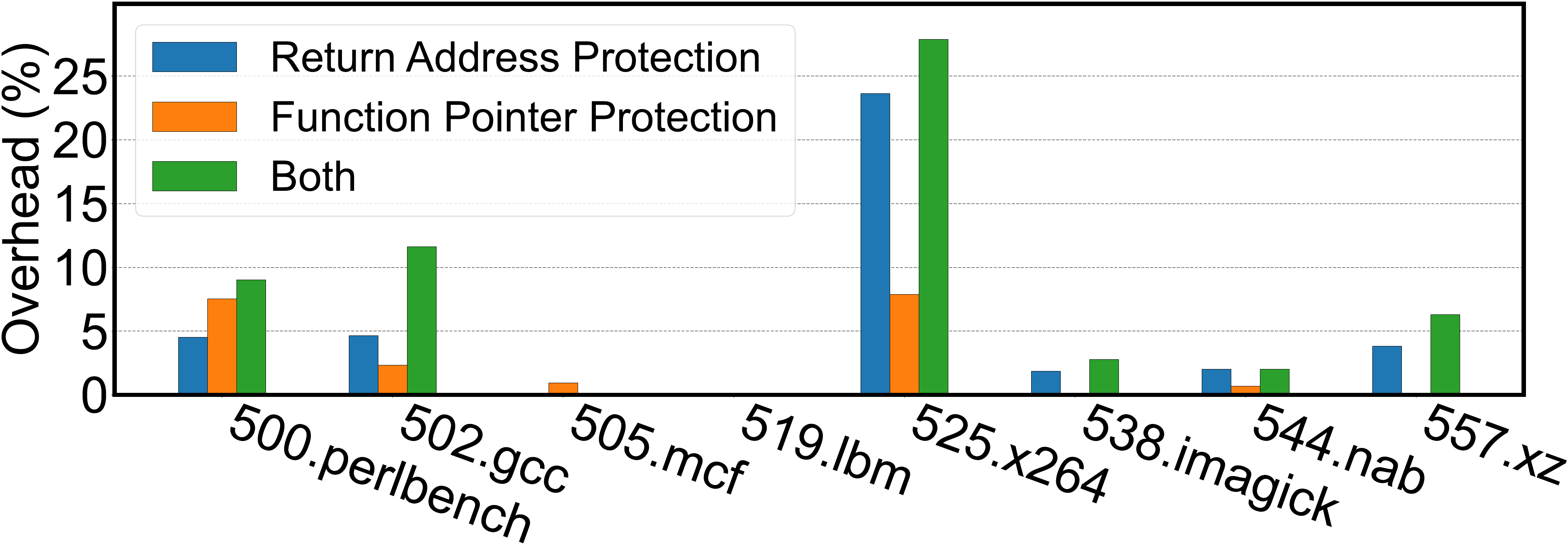}
  \caption{Performance overhead on the SPEC CPU2017 C benchmarks.}
  \label{fig:cpi}
\end{figure}

As illustrated in \autoref{fig:cpi}, protecting return addresses (GCS) incurs a moderate average overhead of 4.83\%. 
Benchmarks with small code footprints, such as \texttt{505.mcf} ($\sim$3K LoCs) and \texttt{519.lbm} ($\sim$1K LoCs), exhibit negligible impact. 
The highest overhead is observed in \texttt{525.x264}, attributed to its video encoding workload which involves deeply nested loops and high-frequency function invocations during macroblock processing.
Protecting function pointers (PIM) results in a lower average overhead of 2.37\% due to their relative infrequency. 
While benchmarks like \texttt{538.imagick} and \texttt{557.xz} show minimal impact, \texttt{perlbench} experiences a higher 7.52\% overhead as its interpreter core relies heavily on function pointers for opcode dispatch. 
Overall, the integrated protection introduces an average overhead of 7.13\%. 
We also tested Nginx and found that CPI alone reduced throughput by just 1.61\%, confirming its practicality in production settings.

\begin{table}[b]
  \caption{lmbench performance overhead ($\mu$s).}
  \label{tab:lmbench}
  \centering
  \small
  \begin{adjustbox}{max width=\linewidth}
  \begin{tabular}{l@{\hspace{5pt}}c@{\hspace{5pt}}c@{\hspace{5pt}}c@{\hspace{5pt}}c@{\hspace{5pt}}c@{\hspace{5pt}}c@{\hspace{5pt}}c@{\hspace{5pt}}c@{\hspace{5pt}}c@{\hspace{5pt}}c}
    \toprule
  \multirow{2}{*}{\textbf{System}} & \textbf{null} & \textbf{null} &  \multirow{2}{*}{\textbf{stat}} & \textbf{open/} & \textbf{select} & \textbf{sig} & \textbf{sig} & \textbf{fork} & \textbf{mmap} & \textbf{page} \\
   & \textbf{call} & \textbf{I/O} & & \textbf{close} & \textbf{TCP} & \textbf{inst} & \textbf{hndl} & \textbf{proc} & \textbf{lat.} & \textbf{fault} \\
    \midrule
    \textbf{Native}   & 0.47 & 0.52 & 1.72 & 2.82 & 6.95 & 0.63 & 2.47 & 1.1K  & 13.2K & 0.81 \\
    \textbf{\sysname} & 3.70 & 4.85 & 8.17 & 11.4 & 11.5 & 3.94 & 7.88 & 12.0K & 49.8K & 1.14 \\
    \bottomrule
  \end{tabular}
  \end{adjustbox}
\end{table}

\noindent \textbf{Runtime Monitoring Impact.}
\sysname intercepts the control flow between a protected process and the OS, introducing additional overhead to kernel services the process relies on. Hooking the trivial \texttt{getpid} call adds about 5,808 cycles (\autoref{tab:operation_cycle}), reflecting only the cost of trap entry/exit, context save-and-restore, and GPT switching that the Monitor performs (\cref{subsec:reduction}). More complex syscalls demand extra work from the Monitor, so we used lmbench v3.0-a9~\cite{2025lmbench} to profile the overall impact. As \autoref{tab:lmbench} shows, monitoring adds less than 10 \textmu s to most operations. The most expensive one is \texttt{mmap}, at 36.6\,ms; this reflects the Monitor validating and updating page tables for the large region that this benchmark repeatedly maps and unmaps, rather than a fixed per-call cost. To characterize the overhead on the path more relevant to application startup, which is dominated by anonymous mappings, we further profiled an allocation-intensive workload that maps 64\,MB of anonymous memory split across $N$ mappings ($N$ from 1 to 8{,}192) and write-touches every page. On the mapping path, \sysname adds about 5.3\,\textmu s per call, with the average per-call cost being 2.1\,\textmu s natively and 7.4\,\textmu s under \sysname. On the first-touch path, \sysname adds about 0.09\,\textmu s per page. Even at the extreme of 8{,}192 mappings, the total startup time is 74\,ms under \sysname against 31\,ms natively.

\subsection{Evaluation on Real-world Applications} \label{subsec:real_world}

\noindent \textbf{Comparison Systems.}
We benchmark against native execution and three alternative schemes: 
(1) Shelter~\cite{zhang2023shelter}, representing process-level isolation where all requests share a single address space; 
(2) Bypass-window-only, our baseline that maps each domain to a bypass window and traps to the Monitor on every switch; 
and (3) lwC~\cite{litton2016light}, adapted from FreeBSD with overheads estimated for privileged emulation~\cite{vahldiek2019erim,gu2022epk}.
We use these case studies primarily to characterize performance and switching behavior under realistic workloads. We do not claim they provide complete end-to-end security against CIVs~\cite{lefeuvre2023assessing} without additional interface hardening.

\begin{figure}[!t]
  \centering
  \includegraphics[width=\linewidth]{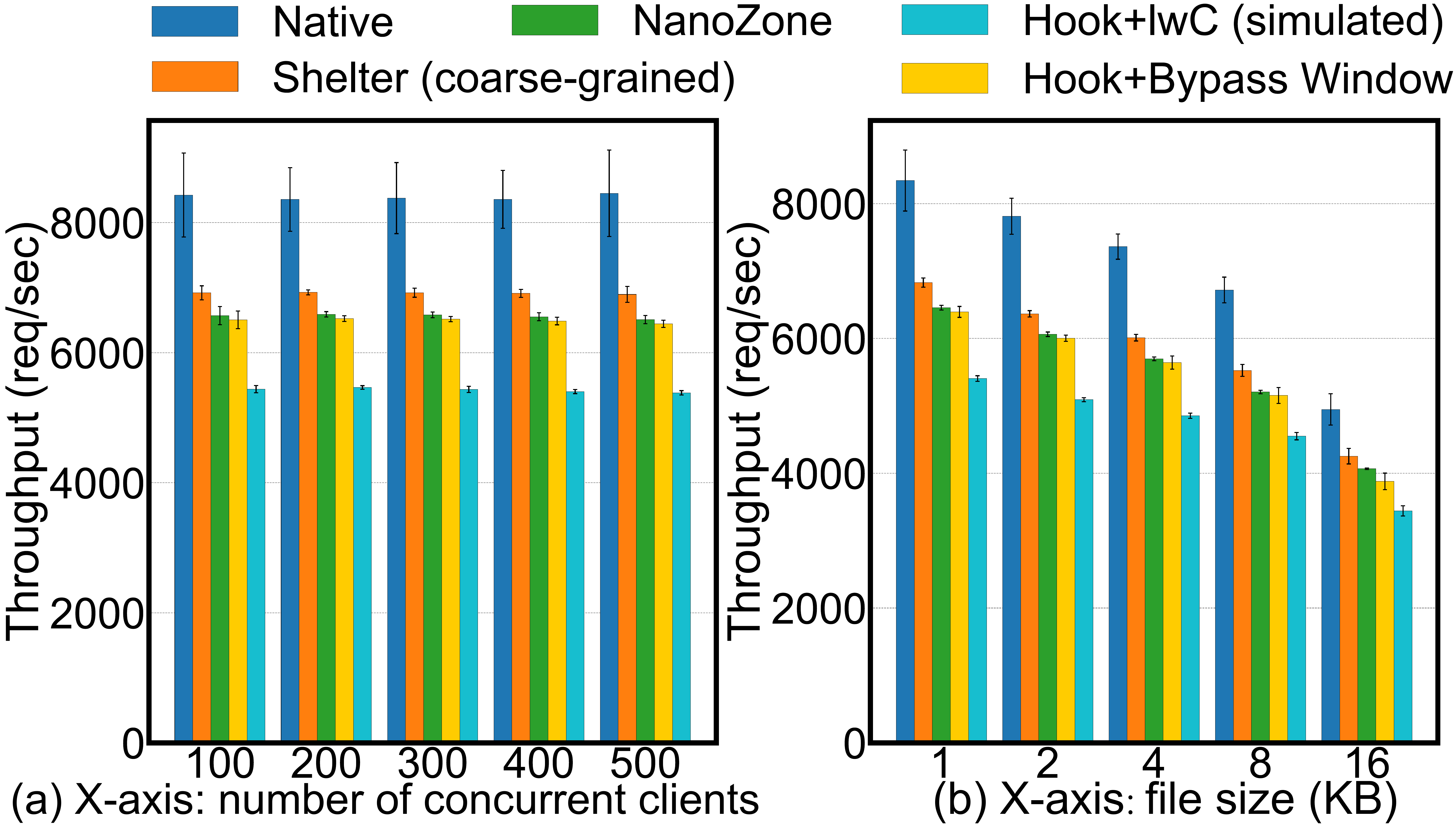}
  \caption{Comparison of Throughput Overhead on Nginx.}
  \label{fig:nginx_combined}
\end{figure}

\noindent \textbf{Case Study 1: Server Application Protection.}
We integrated \sysname into Nginx v1.26.3 linked with OpenSSL v3.3.4~\cite{openssl2024}, isolating each client's SSL session key in a separate domain~\cite{park2019libmpk} while enforcing strict domain-switch validations. 
The server was configured with two worker processes and keep-alive enabled on our dual-core board to handle traffic from a remote \texttt{ab}~\cite{ab2024} client. 
We evaluated throughput across two dimensions: 
(1) \textit{concurrency}, varying clients from 100 to 500 (80 concurrent requests, 22 B payload); and 
(2) \textit{payload size}, ranging from 1 KB to 16 KB (300 clients). 
Reported results represent the average of 50 repetitions under CPU saturation, where any instrumentation overhead directly degrades throughput.

\autoref{fig:nginx_combined}(a) presents the results of the concurrency test. Overall, \sysname incurs a 21.84\% average throughput overhead. Compared to Shelter~\cite{zhang2023shelter}, which provides only privileged isolation, our finer-grained protection adds just 5.15\% overhead, validating the efficiency of our three-tier zone design. Further comparisons reinforce this advantage: relative to the bypass-window-only scheme, \sysname achieves 1.00\% higher throughput, and it outperforms lwC~\cite{litton2016light} by 20.93\%. These improvements stem from the fact that both of those schemes trap to higher privilege levels on each domain switch. \autoref{fig:nginx_combined}(b) shows the file-size test results, which mirror the concurrency findings. Here, the overall average overhead is 21.59\%. Against Shelter, performance drops by 5.12\%, while throughput improves by 1.76\% and 17.70\% over the bypass-window-only scheme and lwC, respectively.

\begin{figure}[!t]
  \centering
  \includegraphics[width=\linewidth]{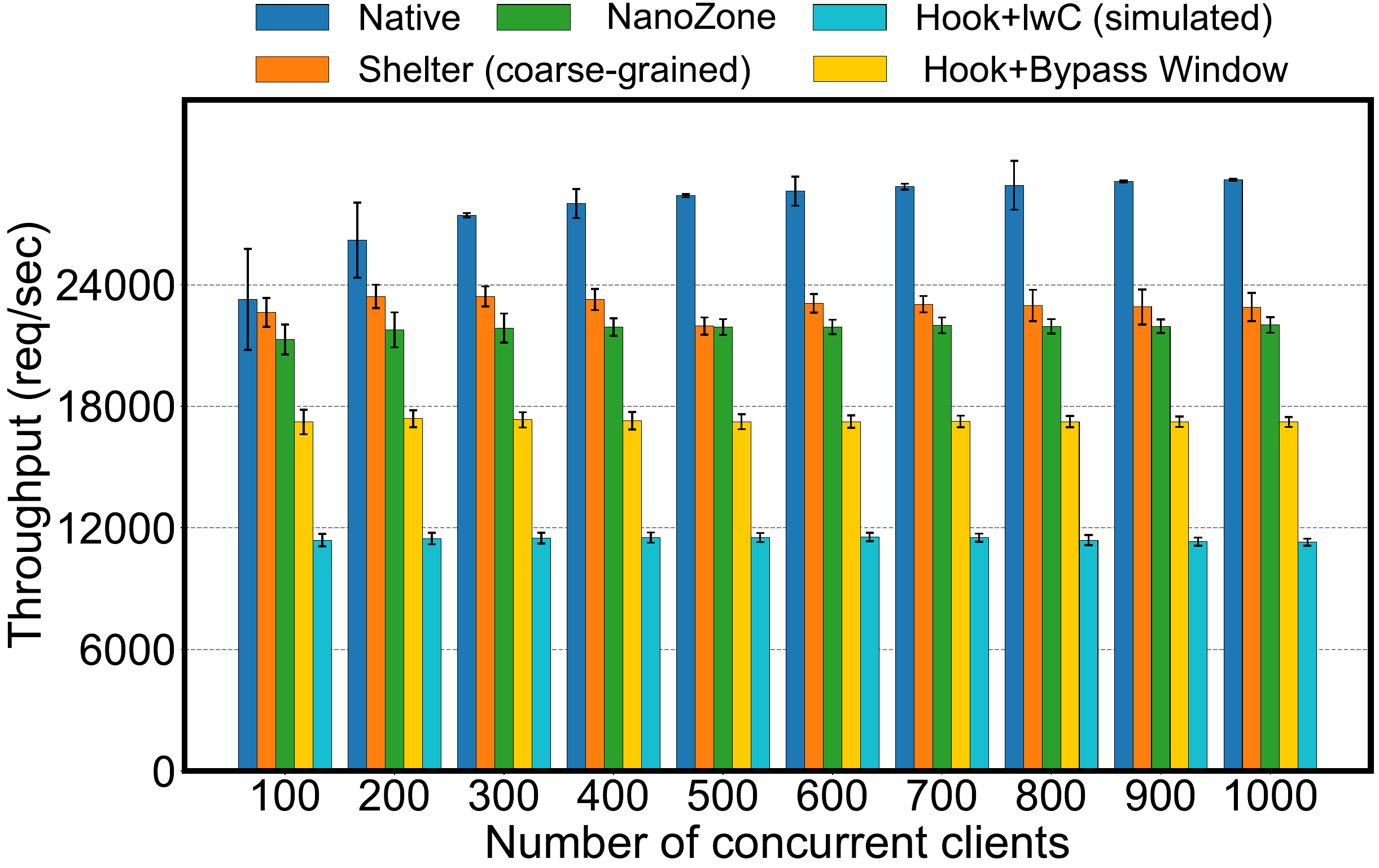}
  \caption{Comparison of Throughput Overhead on Memcached.}
  \label{fig:memcached}
\end{figure}

\begin{figure}[!t]
  \centering
  \includegraphics[width=\linewidth]{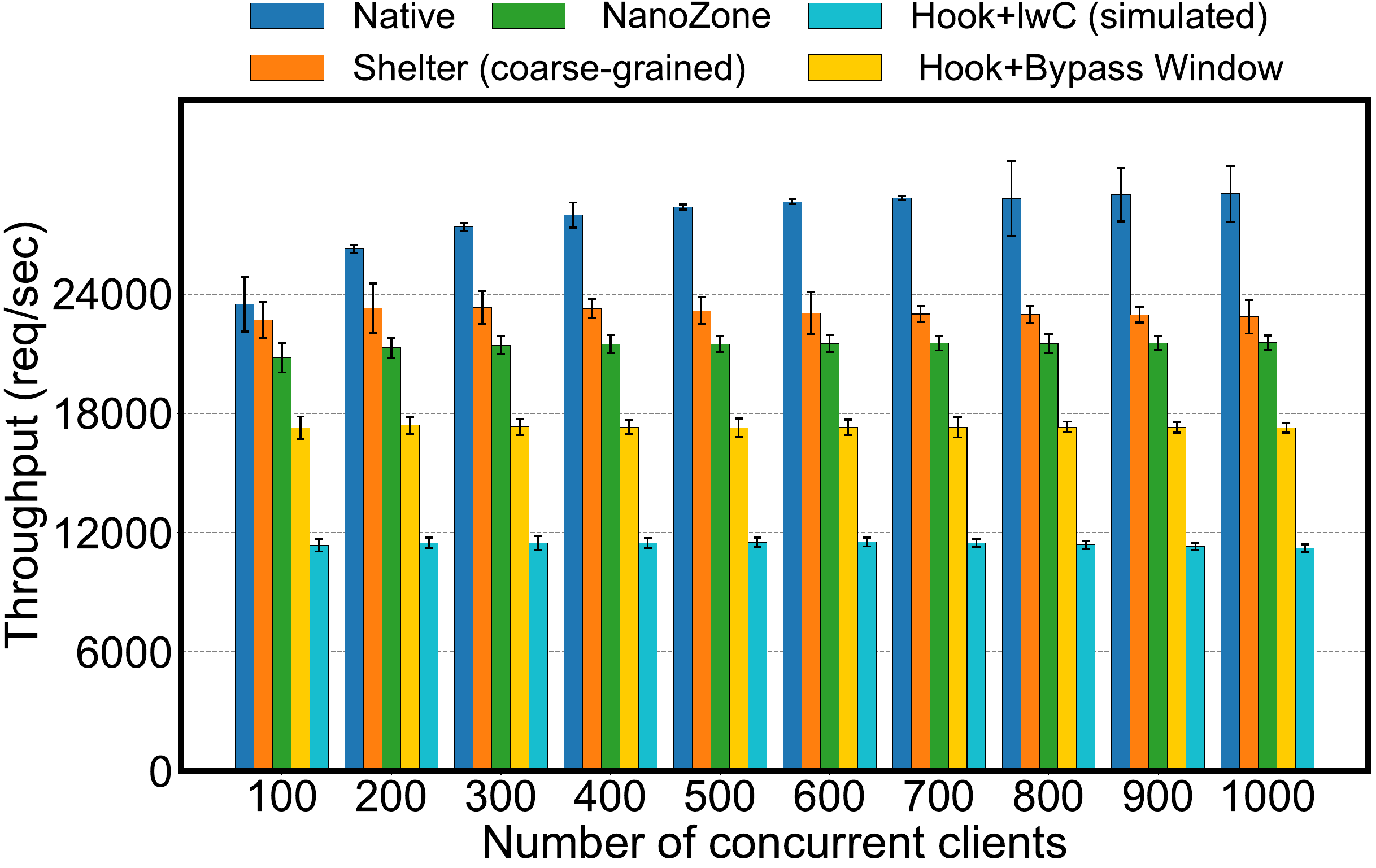}
  \caption{Throughput on Memcached under round-robin domain allocation.}
  \label{fig:memcached_rr}
\end{figure}

\noindent \textbf{Case Study 2: In-Memory Key-Value Shielding.}
We evaluated \sysname on Memcached v1.6.31~\cite{memcached2024} to quantify the overhead of partitioning a high-performance in-memory key-value store. 
The server was configured with a 512 MB preallocated heap, where each client's key-value pairs were isolated in a dedicated domain protected by trampoline-based access controls. 
Running with two worker threads, the server handled traffic generated by a remote \texttt{twemperf}~\cite{twemperf2024} client. 
The workload consisted of 100--1,000 connections, each issuing 30 concurrent requests for 12-byte keys and 64-byte values. 
We report the average throughput computed from 300 repetitions under full CPU saturation.

\begin{figure}[!t]
  \centering
  \includegraphics[width=\linewidth]{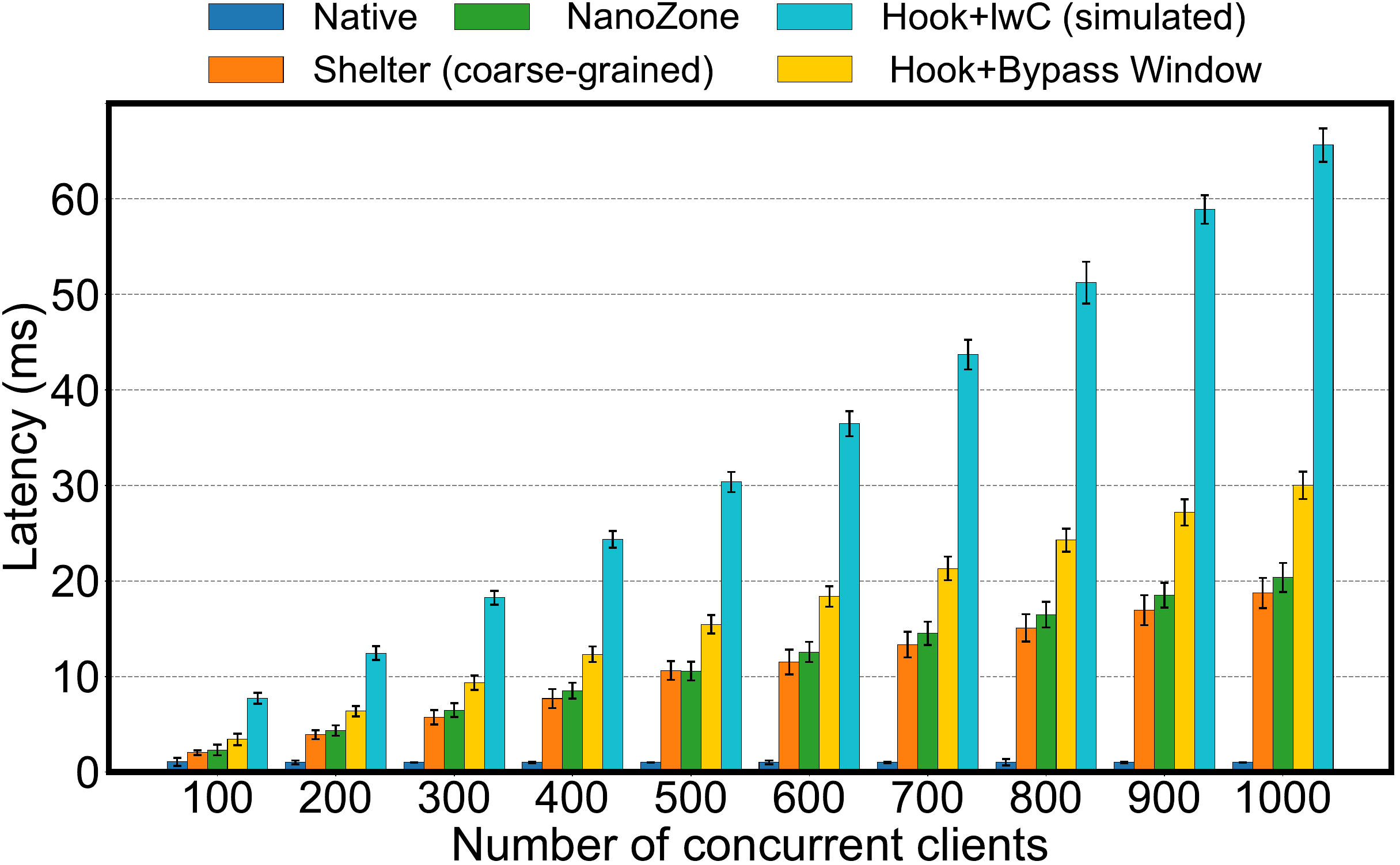}
  \caption{P95 latency on Memcached across concurrency levels.}
  \label{fig:p95}
\end{figure}

\begin{figure}[!t]
  \centering
  \includegraphics[width=\linewidth]{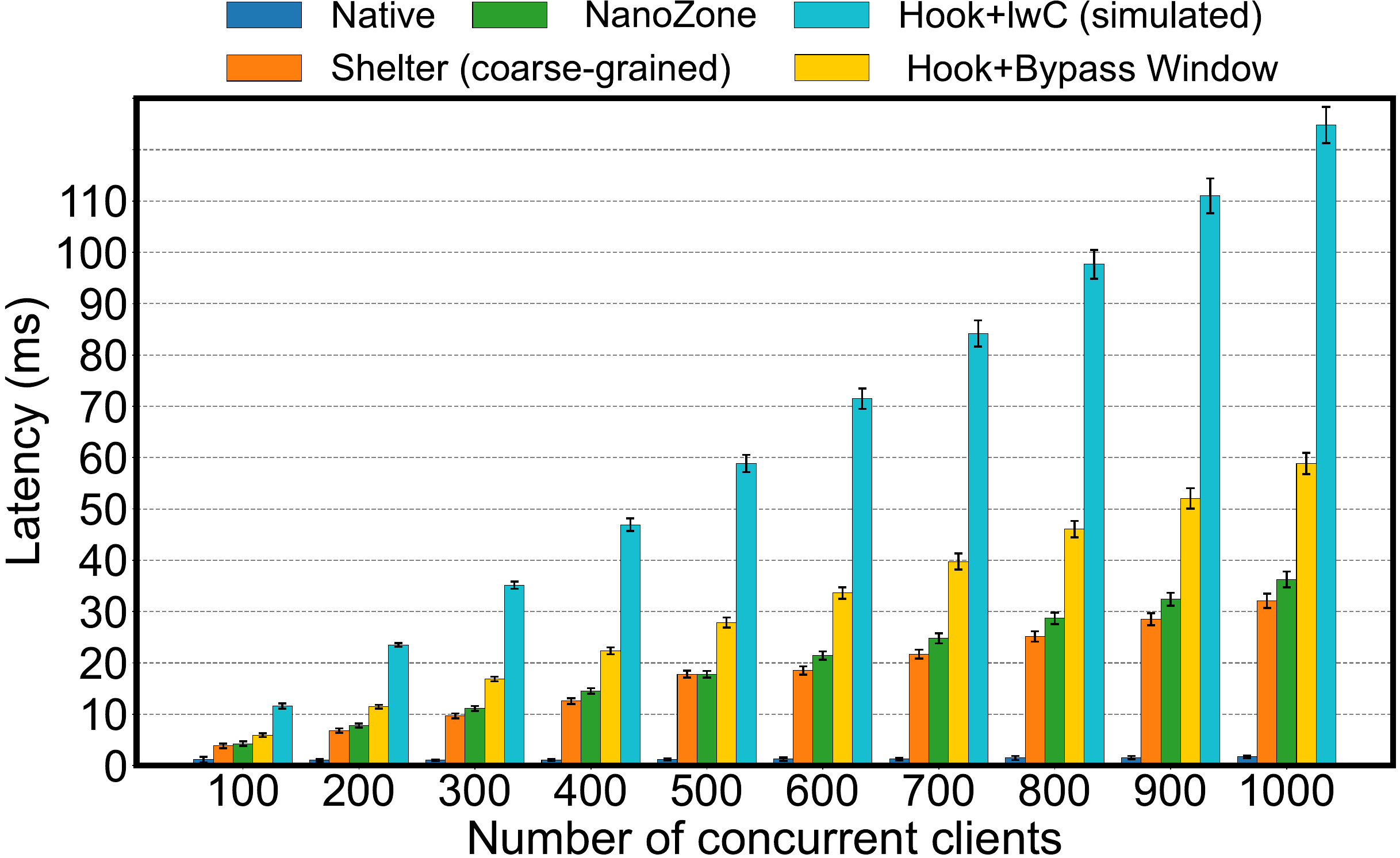}
  \caption{P99 latency on Memcached across concurrency levels.}
  \label{fig:p99}
\end{figure}

\autoref{fig:memcached} illustrates the throughput comparison. \sysname incurs a moderate average overhead of 21.06\% relative to native execution, reaching about 24.6\% under high load. Notably, the majority of this cost is inherent to the baseline privileged isolation (\cref{subsec:reduction}): compared to Shelter~\cite{zhang2023shelter}, \sysname retains 95.23\% of the throughput. In contrast, \sysname significantly outperforms alternative intra-process isolation schemes, achieving 26.60\% and 90.99\% higher throughput than the bypass-window-only baseline and lwC~\cite{litton2016light}, respectively.
These gains validate our domain-allocation optimization (\cref{subsec:extend}), which successfully confines the vast majority of operations to user space, thereby avoiding the costly privileged transitions that penalize the other approaches.

To examine \sysname under frequent privileged switches, we repeated this experiment with round-robin domain allocation, leaving all other settings unchanged. This allocation scatters each thread's domains across L2- and L3-Zones and lowers the L1 hit rate to 77.29\% across 100--1,000 connections. At 1,000 connections, the number of L2 and L3 switches is 6.6$\times$ that at 100 connections. \autoref{fig:memcached_rr} illustrates the resulting throughput. Compared to Shelter, the average drop grows only from 4.77\% under the default allocation to 7.14\%, and is 5.72\% at 1,000 connections. \sysname still outperforms the bypass-window-only baseline at every concurrency level.

\begin{figure*}[!t]
  \centering
  \includegraphics[width=\linewidth]{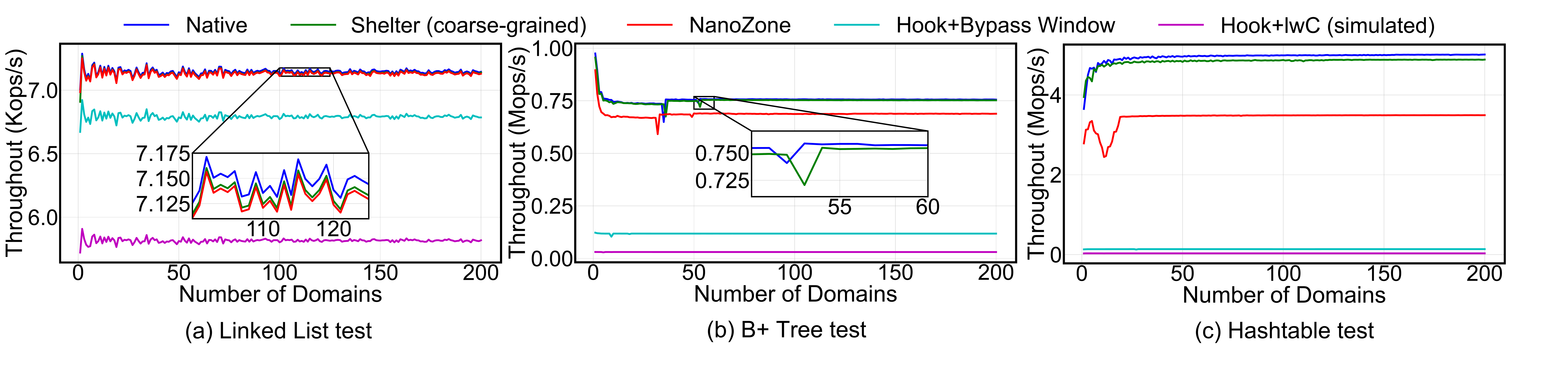}
  \caption{Data-structure benchmarks: (a) linked list (20 KB per domain); (b) B+ tree (5 MB per domain); (c) hash table (20 KB per domain).}
  \label{fig:nvm}
\end{figure*}

We also measured tail latency, which is a primary concern for server workloads. \autoref{fig:p95} and \autoref{fig:p99} report P95 and P99 latency across concurrency levels. Both percentiles grow smoothly and monotonically with load, and \sysname's P99 stays at 1.68$\times$–1.84$\times$ its P95 throughout, indicating a well-behaved tail without sporadic slowdowns. At 1,000 connections, \sysname's P99 is 36.2ms, close to Shelter's 32.1ms and far below the schemes that trap on every switch, whose tails grow much faster (lwC reaches 124.8ms). The absolute values are governed mainly by request queuing under high concurrency, which affects all schemes once throughput saturates; the marginal contribution of \sysname's fine-grained isolation over the coarse-grained Shelter baseline stays within about 4.2 ms across the range.

\noindent \textbf{Case Study 3: NVM Data Isolation.} 
To mitigate risks when mapping NVM directly into a process, we evaluate intra-process isolation using emulated NVM~\cite{xu2020merr,xu2020hardware} and three standard data-structure benchmarks~\cite{gu2022epk} (linked list, B+ tree, hash table). 
We measured throughput by switching domains before and after each random query, varying domain counts from 1 to 200 with over 1,000 repetitions per query.

\autoref{fig:nvm} illustrates that protection overhead is inversely proportional to query latency. 
\sysname incurs negligible overhead (0.21\%) on the compute-intensive linked list, rising to 9.09\% for the B+ tree and 30.68\% for the fast hash table, where domain switching dominates execution time. 
However, \sysname's optimized switching significantly outperforms baselines that rely on privileged traps. 
In the linked-list benchmark, the bypass-window-only scheme and lwC~\cite{litton2016light} incur 5.01\% and 22.61\% higher overhead, respectively. 
These gaps widen dramatically for memory-bound workloads: on the B+ tree, their overheads are 4.82$\times$ and 21.55$\times$ greater than \sysname's, reaching 24.22$\times$ and 108.92$\times$ for the hash table. 
In contrast, Shelter~\cite{zhang2023shelter} maintains near-native performance (max 2.30\% drop) due to low system-call reliance. 
Compared to Shelter, \sysname's fine-grained isolation cost is minimal for slow queries (0.05\% on linked list) but noticeable for faster ones, reaching 8.59\% on the B+ tree and 29.04\% on the hash table.

\section{Discussion \& Limitation} \label{sec:discussion}

\noindent \textbf{Security Boundaries.} \sysname's type-ID mechanism derives a per-function-signature hash via SHAKE-128, enabling fine-grained validation at each indirect call site and substantially shrinking the set of admissible targets compared to coarse-grained mechanisms such as Arm BTI~\cite{arm2023manual}. Same-type reuse remains a residual limitation: an attacker may still redirect execution to a legitimate target with a matching type signature, an inherent constraint of type-based forward-edge CFI. Additionally, microarchitectural side-channel and denial-of-service attacks remain outside \sysname's threat model (\cref{sec:overview}), as in comparable systems, but the tiered design warrants a brief discussion of both. For timing side channels, an L2 and an L3 switch have almost identical cost (\autoref{tab:operation_cycle}), so execution timing separates only a fast L1 switch from an expensive L2/L3 one. This reveals at most whether a switch left the current L2-Zone, not which of the 28 domains in a zone was accessed, a signal far coarser than domain identity. For denial of service, a switch at any tier writes only registers on the executing core, and the global lock is taken when an L3-Zone is allocated or destroyed, never on the switching path. A client issuing frequent expensive switches therefore exercises only the lock-free switching path and cannot block other cores by holding a shared lock. Lock contention is confined to zone allocation and destruction, whose rate is set by connection establishment rather than by request processing. A client also cannot select its own tier, since placement is decided by the Monitor's affinity-based allocation and OS scheduling rather than by the client.
Moreover, \sysname audits the load path through the Monitor, which enforces layout and instruction-level invariants before delegation (\cref{subsec:analysis}); verifying the image content against the tenant's intended binary additionally requires a trusted reference measurement. Full measurement and remote attestation form an orthogonal direction. The Monitor's load-path intervention provides a natural attachment point for such a mechanism, which we leave to future work.

\noindent \textbf{Deployment Boundaries.} Four practical constraints limit immediate deployment. First, \sysname depends on POE/PIE, GCS, and GPC3, which are not yet available on any commercial SoC. Our performance prototype emulates these features on a Rockchip RK3399 board. Native hardware will eliminate the two-stage trap path that dominates our switching estimates and reduce reported overheads accordingly. On transitional hardware that lacks GCS or GPC3, \sysname remains deployable, but with weaker security guarantees. GPC3 can be emulated using a per-PAS-per-GPT scheme, at the cost of losing the fail-closed property provided by our single-GPT design (\cref{subsec:extend}). Similarly, GCS can be replaced with Armv8.3 Pointer Authentication~\cite{liljestrand2019pac,ismail2022tightly}, though this leaves the backward edge vulnerable to pointer-reuse attacks. Second, \sysname's L3-Zone design requires physically contiguous memory of 1\,GB to 64\,GB via the Linux CMA. Even with early boot reservation, large allocations may still fail if pinned or unmovable pages block compaction, or if internal fragmentation leaves no sufficiently large free extent. In that case, the application can be split into multiple processes that each use at most seven domains, so that no L3-Zone is needed. Each switch between these processes, however, passes through the Monitor and costs at least as much as an L2/L3 switch (\autoref{tab:operation_cycle}). The number of processes also grows with the number of principals, and data shared across them must be split or shared through the Monitor. Third, \sysname requires recompilation with a modified LLVM toolchain to instrument function pointers and strip unsafe instructions, which precludes protection of pre-built binaries and introduces a dependency on a trusted build environment. Fourth, code pages are fixed at launch. After delegation, the Monitor denies any transition of a page to executable (\cref{subsec:reduction}). Runtime code loading such as \texttt{dlopen} or JIT compilation is therefore unsupported. This is a deliberate trade-off that closes the runtime code-injection surface. Our evaluated server workloads load all modules at startup and are unaffected.

\noindent \textbf{Performance Boundaries.} \sysname's low overhead relies on request locality keeping the majority of domain switches in the user-space L1 tier. For workloads whose access pattern lacks such locality, where consecutive operations scatter across many domains, the share of privileged L2 and L3 switches rises and the average switching cost increases accordingly (see \cref{subsec:real_world} for a round-robin stress test); the locality-aware allocation (\cref{subsec:extend}) is therefore most effective when requests exhibit the temporal locality typical of per-connection server workloads. 
Furthermore, \sysname intercepts memory-mapping operations, so workloads dominated by repeated mapping and unmapping of large shared regions incur higher Monitor overhead (\cref{subsec:microbenchmarks}). It is therefore best suited to server applications whose bottleneck lies in compute or network I/O. Allocation-intensive startup nonetheless remains bounded (\cref{subsec:microbenchmarks}), and pre-allocating memory pools via \texttt{zone\_alloc} further reduces mapping frequency.

\section{Related Work} \label{sec:related_work}

\noindent \textbf{MPK-based Mechanisms.} Intel MPK has become the standard primitive for fast intra-process isolation on x86, and has been used to secure language runtimes~\cite{kirth2022pkru,lei2023put}, compartmentalize OS components~\cite{lefeuvre2022flexos,lu2023moat}, and optimize process abstractions~\cite{lin2024fast,vahldiek2019erim}. A separate line of work scales MPK beyond its domain limit, by modifying hardware, evicting and reusing domain IDs, or reusing IDs through virtualization-level page tables (\cref{subsec:motivation}). \sysname differs in two respects. First, it reuses POE/PIE indexes across distinct physical address spaces through GPT, rather than through extra tables or ID eviction. Second, unlike schemes that depend on a trusted OS or hypervisor to manage these structures, it anchors enforcement in the root-world Monitor and stays secure under an adversarial kernel. More broadly, these advances rely on x86-specific hardware, whereas \sysname delivers an equally lightweight facility on Arm.

\noindent \textbf{Hardware-assisted Control-flow Integrity.} While x86 leverages Intel CET for native shadow stacks~\cite{maar2024beyond}, Arm architectures long lacked an equivalent hardware mechanism. Prior works approximated this functionality using Pointer Authentication (e.g., PARTS~\cite{liljestrand2019pac}, PACTight~\cite{ismail2022tightly}) or PAN extensions (e.g., PANIC~\cite{xu2023panic}), but these approaches remain susceptible to pointer reuse or incur substantial complexity. With the advent of Armv9, \sysname is the first to leverage the native Guarded Control Stack (GCS) to enforce strict, bidirectional control-flow integrity.

\noindent \textbf{OS Security Monitors.} A range of systems interpose a trusted layer at a higher privilege level to protect applications from an untrusted OS. Hypervisor-based designs such as Overshadow~\cite{chen2008overshadow} and InkTag~\cite{hofmann2013inktag} shield an application's memory from the OS. However, they protect the application as a single unit, leaving its interior exposed, and add a full hypervisor to the TCB. Others reduce the trusted stack using hardware modifications~\cite{song2024interstellar} or platform-specific mechanisms such as AMD VMPLs~\cite{wang2024road,ahmad2023veil}. For designs built on Arm CCA primitives, placing the monitor at Realm EL2 is suboptimal because it pulls the complex RMM into the TCB. \sysname departs from these in two ways. First, it enforces a fine-grained boundary within the process, closing the intra-process gap these systems leave open. Second, it anchors security in a minimal root-world monitor at EL3, rather than trusting a hypervisor or the RMM. Unlike RContainer~\cite{zhourcontainer}, which offloads some tasks to a mini-OS to reduce root-world usage, \sysname consolidates critical enforcement in the monitor to maintain a single, minimal root of trust.

\noindent\textbf{CCA-based Isolation.} Shelter~\cite{zhang2023shelter} provides process-level CCA isolation, but treats each application as a single protection unit. As a result, it provides no intra-process separation and leaves the application interior unprotected. \sysname adds this interior boundary at a small marginal cost, only 5.2\% over Shelter (\cref{subsec:real_world}). Other systems apply CCA primitives in different directions. CAGE~\cite{wang2024cage} and Portal~\cite{sang2025portal} use them for confidential GPU computing and secure device access. Closest to our work is HiveTEE~\cite{huang2026hivetee}, a concurrent system that, like \sysname, reuses GPTs to scale domains beyond a fixed hardware limit, virtualizing MTE tags where \sysname virtualizes POE/PIE indexes. The key difference is the switching path: every HiveTEE domain entry is mediated by the Monitor, whereas \sysname performs most switches in user space (\autoref{tab:switch_cycle}). Because these user-space gates grant permissions to the calling thread, \sysname must enforce CPI on the path leading to them (\cref{subsec:cfi}) and shield the entire process from the OS. HiveTEE validates entries but leaves CFI to orthogonal work, and shields only annotated functions, which keeps its overhead below 3\% but leaves the rest of the application untrusted. HiveTEE's MTE tags, in turn, offer 16-byte granularity but must be forced on every memory access through instrumentation.

\section{Conclusion} \label{sec:conclusion}

Arm CCA isolates software only at VM granularity, leaving applications exposed to a compromised OS and to vulnerabilities in their own code.
\sysname addresses both by protecting normal-world processes from a minimal root-world monitor and partitioning them into fine-grained domains.
It combines POE, PIE, and GPT into a three-tier hierarchy, keeps most domain switches in user space through locality-aware allocation, and protects the switch paths with CPI.
Across our server case studies, this fine-grained protection costs about 5\% over process-level isolation.

\bibliographystyle{IEEEtran}
\bibliography{main}

@inproceedings{durumeric2014matter,
  title={The Matter of {Heartbleed}},
  author={Durumeric, Zakir and Li, Frank and Kasten, James and Amann, Johanna and Beekman, Jethro and Payer, Mathias and Weaver, Nicolas and Adrian, David and Paxson, Vern and Bailey, Michael and others},
  booktitle={Proceedings of the 2014 conference on internet measurement conference},
  pages={475--488},
  year={2014}
}

@inproceedings{gu2022hardware,
  title={A {Hardware-Software} co-design for efficient {Intra-Enclave} isolation},
  author={Gu, Jinyu and Zhu, Bojun and Li, Mingyu and Li, Wentai and Xia, Yubin and Chen, Haibo},
  booktitle={31st USENIX Security Symposium (USENIX Security 22)},
  pages={3129--3145},
  year={2022}
}

@inproceedings{park2020nested,
  title={Nested Enclave: Supporting Fine-grained Hierarchical Isolation with {SGX}},
  author={Park, Joongun and Kang, Naegyeong and Kim, Taehoon and Kwon, Youngjin and Huh, Jaehyuk},
  booktitle={2020 ACM/IEEE 47th Annual International Symposium on Computer Architecture (ISCA)},
  pages={776--789},
  year={2020},
  organization={IEEE}
}

@inproceedings{voulimeneas2022you,
  title={You shall not (by)pass!: practical, secure, and fast {PKU-based} sandboxing},
  author={Voulimeneas, Alexios and Vinck, Jonas and Mechelinck, Ruben and Volckaert, Stijn},
  booktitle={Proceedings of the Seventeenth European Conference on Computer Systems},
  pages={266--282},
  year={2022}
}

@inproceedings{schrammel2020donky,
  title={Donky: Domain Keys {\textendash} Efficient {In-Process} Isolation for {RISC-V} and x86},
  author={Schrammel, David and Weiser, Samuel and Steinegger, Stefan and Schwarzl, Martin and Schwarz, Michael and Mangard, Stefan and Gruss, Daniel},
  booktitle={29th USENIX Security Symposium (USENIX Security 20)},
  pages={1677--1694},
  year={2020}
}

@inproceedings{park2019libmpk,
  title={libmpk: Software Abstraction for {Intel} Memory Protection Keys ({Intel} {MPK})},
  author={Park, Soyeon and Lee, Sangho and Xu, Wen and Moon, Hyungon and Kim, Taesoo},
  booktitle={2019 USENIX Annual Technical Conference (USENIX ATC 19)},
  pages={241--254},
  year={2019}
}

@inproceedings{vahldiek2019erim,
  title={{ERIM}: Secure, Efficient In-process Isolation with Protection Keys ({MPK})},
  author={Vahldiek-Oberwagner, Anjo and Elnikety, Eslam and Duarte, Nuno O and Sammler, Michael and Druschel, Peter and Garg, Deepak},
  booktitle={28th USENIX Security Symposium (USENIX Security 19)},
  pages={1221--1238},
  year={2019}
}

@inproceedings{xu2023panic,
  title={{PANIC}: {PAN-assisted} Intra-process Memory Isolation on {ARM}},
  author={Xu, Jiali and Xie, Mengyao and Wu, Chenggang and Zhang, Yinqian and Li, Qijing and Huang, Xuan and Lai, Yuanming and Kang, Yan and Wang, Wei and Wei, Qiang and others},
  booktitle={Proceedings of the 2023 ACM SIGSAC Conference on Computer and Communications Security},
  pages={919--933},
  year={2023}
}

@inproceedings{dinh2023capacity,
  title={Capacity: Cryptographically-Enforced In-Process Capabilities for Modern {ARM} Architectures},
  author={Dinh Duy, Kha and Cho, Kyuwon and Noh, Taehyun and Lee, Hojoon},
  booktitle={Proceedings of the 2023 ACM SIGSAC Conference on Computer and Communications Security},
  pages={874--888},
  year={2023}
}

@misc{arm2023manual,
  title={Arm® Architecture Reference Manual for {A-profile} architecture},
  author={Arm},
  year={2024},
  note={\url{https://developer.arm.com/documentation/ddi0487/latest/}}
}

@misc{intel2023shadow,
  title={Complex Shadow-Stack Updates ({Intel} Control-Flow Enforcement Technology)},
  author={Intel},
  year={2023},
  note={\url{https://www.intel.com/content/www/us/en/content-details/785687/complex-shadow-stack-updates-intel-control-flow-enforcement-technology.html}}
}

@misc{arm2023cca,
  title={Confidential Compute Architecture},
  author={Arm},
  year={2023},
  note={\url{https://www.arm.com/architecture/security-features/arm-confidential-compute-architecture}}
}

@inproceedings{zhang2023shelter,
  title={{SHELTER}: Extending {Arm} {CCA} with Isolation in User Space},
  author={Zhang, Yiming and Hu, Yuxin and Ning, Zhenyu and Zhang, Fengwei and Luo, Xiapu and Huang, Haoyang and Yan, Shoumeng and He, Zhengyu},
  booktitle={32nd USENIX Security Symposium (USENIX Security 23)},
  year={2023},
  pages={6257--6274}
}

@inproceedings{hofmann2013inktag,
  title={{InkTag}: secure applications on an untrusted operating system},
  author={Hofmann, Owen S and Kim, Sangman and Dunn, Alan M and Lee, Michael Z and Witchel, Emmett},
  booktitle={Proceedings of the eighteenth international conference on Architectural support for programming languages and operating systems},
  pages={265--278},
  year={2013}
}

@misc{arm2023fvp,
  title={Fixed Virtual Platforms},
  author={Arm},
  year={2023},
  note={\url{https://support.arm.com/tools-and-software/fixed-virtual-platforms}}
}

@inproceedings{lee2020off,
  title={An {Off-Chip} attack on hardware enclaves via the memory bus},
  author={Lee, Dayeol and Jung, Dongha and Fang, Ian T and Tsai, Chia-Che and Popa, Raluca Ada},
  booktitle={29th USENIX Security Symposium (USENIX Security 20)},
  year={2020},
  pages={487--504}
}

@inproceedings{yitbarek2017cold,
  title={Cold boot attacks are still hot: Security analysis of memory scramblers in modern processors},
  author={Yitbarek, Salessawi Ferede and Aga, Misiker Tadesse and Das, Reetuparna and Austin, Todd},
  booktitle={2017 IEEE International Symposium on High Performance Computer Architecture (HPCA)},
  pages={313--324},
  year={2017},
  organization={IEEE}
}

@article{kim2014flipping,
  title={Flipping bits in memory without accessing them: An experimental study of {DRAM} disturbance errors},
  author={Kim, Yoongu and Daly, Ross and Kim, Jeremie and Fallin, Chris and Lee, Ji Hye and Lee, Donghyuk and Wilkerson, Chris and Lai, Konrad and Mutlu, Onur},
  journal={ACM SIGARCH Computer Architecture News},
  volume={42},
  number={3},
  pages={361--372},
  year={2014},
  publisher={ACM New York, NY, USA}
}

@article{lipp2018meltdown,
  title={Meltdown: reading kernel memory from user space},
  author={Lipp, Moritz and Schwarz, Michael and Gruss, Daniel and Prescher, Thomas and Haas, Werner and Horn, Jann and Mangard, Stefan and Kocher, Paul and Genkin, Daniel and Yarom, Yuval and Hamburg, Mike and Strackx, Raoul},
  journal={Communications of the ACM},
  volume={63},
  number={6},
  pages={46--56},
  year={2020}
}

@article{kocher2020spectre,
  title={Spectre attacks: exploiting speculative execution},
  author={Kocher, Paul and Horn, Jann and Fogh, Anders and Genkin, Daniel and Gruss, Daniel and Haas, Werner and Hamburg, Mike and Lipp, Moritz and Mangard, Stefan and Prescher, Thomas and others},
  journal={Communications of the ACM},
  volume={63},
  number={7},
  pages={93--101},
  year={2020},
  publisher={ACM New York, NY, USA}
}

@misc{arm2023introcca,
  title={Learn the architecture - {Realm} Management Extension Guide},
  author={Arm},
  year={2023},
  note={\url{https://developer.arm.com/documentation/den0126/latest}}
}

@inproceedings{shacham2007geometry,
  title={The geometry of innocent flesh on the bone: return-into-libc without function calls (on the x86)},
  author={Shacham, Hovav},
  booktitle={Proceedings of the 14th ACM conference on Computer and communications security},
  pages={552--561},
  year={2007}
}

@inproceedings{bletsch2011jump,
  title={Jump-oriented programming: a new class of code-reuse attack},
  author={Bletsch, Tyler and Jiang, Xuxian and Freeh, Vince W and Liang, Zhenkai},
  booktitle={Proceedings of the 6th ACM symposium on information, computer and communications security},
  pages={30--40},
  year={2011}
}

@misc{intel2023tdx,
  title={Intel Trust Domain Extensions ({Intel} {TDX})},
  author={Intel},
  year={2023},
  note={\url{https://www.intel.com/content/www/us/en/developer/articles/technical/intel-trust-domain-extensions.html}}
}

@misc{amd2023sev,
  title={{AMD} Secure Encrypted Virtualization ({SEV})},
  author={AMD},
  year={2023},
  note={\url{https://www.amd.com/en/developer/sev.html}}
}

@misc{intel2023sgx,
  title={Intel® Software Guard Extensions ({Intel}® {SGX})},
  author={Intel},
  year={2023},
  note={\url{https://www.intel.com/content/www/us/en/products/details/processors/xeon/features/software-guard-extensions.html}}
}

@article{checkoway2013iago,
  title={Iago attacks: why the system call {API} is a bad untrusted {RPC} interface},
  author={Checkoway, Stephen and Shacham, Hovav},
  journal={ACM SIGARCH Computer Architecture News},
  volume={41},
  number={1},
  pages={253--264},
  year={2013},
  publisher={ACM New York, NY, USA}
}

@article{chen2008overshadow,
  title={Overshadow: a virtualization-based approach to retrofitting protection in commodity operating systems},
  author={Chen, Xiaoxin and Garfinkel, Tal and Lewis, E Christopher and Subrahmanyam, Pratap and Waldspurger, Carl A and Boneh, Dan and Dwoskin, Jeffrey and Ports, Dan RK},
  journal={ACM SIGOPS Operating Systems Review},
  volume={42},
  number={2},
  pages={2--13},
  year={2008},
  publisher={ACM New York, NY, USA}
}

@inproceedings{zeng2023retspill,
  title={{RetSpill}: Igniting User-Controlled Data to Burn Away {Linux} Kernel Protections},
  author={Zeng, Kyle and Lin, Zhenpeng and Lu, Kangjie and Xing, Xinyu and Wang, Ruoyu and Doup{\'e}, Adam and Shoshitaishvili, Yan and Bao, Tiffany},
  booktitle={Proceedings of the 2023 ACM SIGSAC Conference on Computer and Communications Security},
  pages={3093--3107},
  year={2023}
}

@inproceedings{lin2022dirtycred,
  title={{DirtyCred}: Escalating Privilege in {Linux} Kernel},
  author={Lin, Zhenpeng and Wu, Yuhang and Xing, Xinyu},
  booktitle={Proceedings of the 2022 ACM SIGSAC Conference on Computer and Communications Security},
  pages={1963--1976},
  year={2022}
}

@inproceedings{gu2022epk,
  title={{EPK}: Scalable and Efficient Memory Protection Keys},
  author={Gu, Jinyu and Li, Hao and Li, Wentai and Xia, Yubin and Chen, Haibo},
  booktitle={2022 USENIX Annual Technical Conference (USENIX ATC 22)},
  pages={609--624},
  year={2022}
}

@inproceedings{wang2024cage,
title={{CAGE}: Complementing {Arm} {CCA} with {GPU} Extensions},
author={Wang, Chenxu and Zhang, Fengwei and Deng, Yunjie and Leach, Kevin and Cao, Jiannong and Ning, Zhenyu and Yan, Shoumeng and He, Zhengyu},
booktitle={Proceedings of the 31st Annual Network and Distributed System Security Symposium},
year={2024}
}

@inproceedings{zhou2014armlock,
  title={{ARMlock}: Hardware-based Fault Isolation for {ARM}},
  author={Zhou, Yajin and Wang, Xiaoguang and Chen, Yue and Wang, Zhi},
  booktitle={Proceedings of the 2014 ACM SIGSAC conference on computer and communications security},
  pages={558--569},
  year={2014}
}

@misc{arm2024pio,
  title={Permission indirection and permission overlay extensions},
  author={Arm},
  year={2024},
  note={\url{https://developer.arm.com/documentation/102376/0200/Permission-indirection-and-permission-overlay-extensions}}
}

@misc{intel2024mpk,
  title={Intel® 64 and {IA-32} Architectures Software Developer Manuals},
  author={Intel},
  year={2024},
  note={\url{https://www.intel.com/content/www/us/en/developer/articles/technical/intel-sdm.html}}
}

@misc{ab2024,
  title={ab - {Apache} {HTTP} server benchmarking tool},
  author={{Apache Software Foundation}},
  year={2024},
  note={\url{https://httpd.apache.org/docs/current/programs/ab.html}}
}

@inproceedings{chen2016shreds,
  title={Shreds: Fine-grained execution units with private memory},
  author={Chen, Yaohui and Reymondjohnson, Sebassujeen and Sun, Zhichuang and Lu, Long},
  booktitle={2016 IEEE Symposium on Security and Privacy (SP)},
  pages={56--71},
  year={2016},
  organization={IEEE}
}

@inproceedings{kirth2022pkru,
  title={{PKRU-safe}: automatically locking down the heap between safe and unsafe languages},
  author={Kirth, Paul and Dickerson, Mitchel and Crane, Stephen and Larsen, Per and Dabrowski, Adrian and Gens, David and Na, Yeoul and Volckaert, Stijn and Franz, Michael},
  booktitle={Proceedings of the Seventeenth European Conference on Computer Systems},
  pages={132--148},
  year={2022}
}

@inproceedings{liu2015thwarting,
  title={Thwarting memory disclosure with efficient hypervisor-enforced intra-domain isolation},
  author={Liu, Yutao and Zhou, Tianyu and Chen, Kexin and Chen, Haibo and Xia, Yubin},
  booktitle={Proceedings of the 22nd ACM SIGSAC Conference on Computer and Communications Security},
  pages={1607--1619},
  year={2015}
}

@inproceedings{litton2016light,
  title={{Light-Weight} Contexts: An {OS} Abstraction for Safety and Performance},
  author={Litton, James and Vahldiek-Oberwagner, Anjo and Elnikety, Eslam and Garg, Deepak and Bhattacharjee, Bobby and Druschel, Peter},
  booktitle={12th USENIX Symposium on Operating Systems Design and Implementation (OSDI 16)},
  pages={49--64},
  year={2016}
}

@inproceedings{yuan2024lightzone,
  title={{LightZone}: Lightweight Hardware-Assisted In-Process Isolation for {ARM64}},
  author={Yuan, Ziqi and Hong, Siyu and Guo, Ruorong and Chang, Rui and Gao, Mingyu and Shen, Wenbo and Zhou, Yajin},
  booktitle={Proceedings of the 25th International Middleware Conference},
  pages={467--480},
  year={2024}
}

@misc{bypass2024arm,
  title={Granule Protection Check Bypass Window Register ({EL3})},
  author={Arm},
  year={2024},
  note={\url{https://developer.arm.com/documentation/ddi0601/2024-12/AArch64-Registers/GPCBW-EL3--Granule-Protection-Check-Bypass-Window-Register--EL3-}}
}

@inproceedings{lefeuvre2024sok,
  title={{SoK}: Software Compartmentalization},
  author={Lefeuvre, Hugo and Dautenhahn, Nathan and Chisnall, David and Olivier, Pierre},
  booktitle={2025 IEEE Symposium on Security and Privacy (SP)},
  pages={3107--3126},
  year={2025},
  organization={IEEE}
}

@inproceedings{kuvaiskii2024gramine,
  title={{Gramine-TDX}: A Lightweight {OS} Kernel for Confidential {VMs}},
  author={Kuvaiskii, Dmitrii and Stavrakakis, Dimitrios and Qin, Kailun and Xing, Cedric and Bhatotia, Pramod and Vij, Mona},
  booktitle={Proceedings of the 2024 on ACM SIGSAC Conference on Computer and Communications Security},
  pages={4598--4612},
  year={2024}
}

@inproceedings{zhourcontainer,
  title={{RContainer}: A Secure Container Architecture through Extending {ARM} {CCA} Hardware Primitives},
  author={Zhou, Qihang and Cao, Wenzhuo and Jia, Xiaoqi and Liu, Peng and Zhang, Shengzhi and Chen, Jiayun and Xu, Shaowen and Song, Zhenyu},
  booktitle={NDSS},
  year={2025}
}

@inproceedings{wang2024road,
  title={The Road to Trust: Building Enclaves within Confidential {VMs}},
  author={Wang, Wenhao and Song, Linke and Mei, Benshan and Liu, Shuang and Zhao, Shijun and Yan, Shoumeng and Wang, XiaoFeng and Meng, Dan and Hou, Rui},
  booktitle={NDSS},
  year={2025}
}

@inproceedings{liljestrand2019pac,
  title={{PAC} it up: Towards pointer integrity using {ARM} pointer authentication},
  author={Liljestrand, Hans and Nyman, Thomas and Wang, Kui and Perez, Carlos Chinea and Ekberg, Jan-Erik and Asokan, N},
  booktitle={28th USENIX Security Symposium (USENIX Security 19)},
  pages={177--194},
  year={2019}
}

@inproceedings{sang2025portal,
  title={Portal: Fast and Secure Device Access with {Arm} {CCA} for Modern {Arm} Mobile System-on-Chips ({SoCs})},
  author={Sang, Fan and Lee, Jaehyuk and Zhang, Xiaokuan and Kim, Taesoo},
  booktitle={2025 IEEE Symposium on Security and Privacy (SP)},
  pages={4099--4116},
  year={2025},
  organization={IEEE}
}

@article{huang2026hivetee,
  title={{HiveTEE}: Scalable and Fine-Grained Isolated Domains With {RME} and {MTE} Co-Assisted},
  author={Huang, Haoyang and Zhang, Fengwei},
  journal={IEEE Transactions on Information Forensics and Security},
  volume={21},
  pages={2300--2312},
  year={2026},
  doi={10.1109/TIFS.2026.3660599},
  publisher={IEEE}
}

@inproceedings{ismail2022tightly,
  title={Tightly seal your sensitive pointers with {PACTight}},
  author={Ismail, Mohannad and Quach, Andrew and Jelesnianski, Christopher and Jang, Yeongjin and Min, Changwoo},
  booktitle={31st USENIX Security Symposium (USENIX Security 22)},
  pages={3717--3734},
  year={2022}
}

@misc{2024armx925,
  title={{Arm} {Cortex-X925} Core Technical Reference Manual},
  author={Arm},
  year={2024},
  note={\url{https://developer.arm.com/documentation/102807/latest/}}
}

@misc{2024radxarockpi,
  title={{ROCK} {4B}},
  author={{Radxa}},
  year={2024},
  note={\url{https://radxa.com/products/rock4/4b/}}
}

@misc{2024linuxpatchwork,
  title={Patchwork - The {Linux} Kernel Archives},
  author={{Linux Foundation}},
  year={2024},
  note={\url{https://patchwork.kernel.org/}}
}

@misc{2025speccpu2017,
  title={{SPEC} {CPU}® 2017 benchmark},
  author={{Standard Performance Evaluation Corporation}},
  year={2025},
  note={\url{https://www.spec.org/cpu2017/}}
}

@misc{2025lmbench,
  title={{LMbench} - Tools for Performance Analysis},
  author={McVoy, Larry and Staelin, Carl},
  year={2025},
  note={\url{https://lmbench.sourceforge.net/}}
}

@misc{openssl2024,
  title={{OpenSSL} Library},
  author={{OpenSSL Foundation}},
  year={2024},
  note={\url{https://www.openssl.org/}}
}

@misc{memcached2024,
  title={memcached - a distributed memory object caching system},
  author={Vorobey, Anatoly and Fitzpatrick, Brad and Kasindorf, Alan "Dormando"},
  year={2024},
  note={\url{https://memcached.org/}}
}

@misc{twemperf2024,
  title={A tool for measuring memcached server performance},
  author={{Twitter}},
  year={2024},
  note={\url{https://github.com/twitter-archive/twemperf/}}
}

@inproceedings{xu2020merr,
  title={{MERR}: Improving Security of Persistent Memory Objects via Efficient Memory Exposure Reduction and Randomization},
  author={Xu, Yuanchao and Solihin, Yan and Shen, Xipeng},
  booktitle={Proceedings of the Twenty-Fifth International Conference on Architectural Support for Programming Languages and Operating Systems},
  pages={987--1000},
  year={2020}
}

@inproceedings{xu2020hardware,
  title={Hardware-based domain virtualization for intra-process isolation of persistent memory objects},
  author={Xu, Yuanchao and Ye, ChenCheng and Solihin, Yan and Shen, Xipeng},
  booktitle={2020 ACM/IEEE 47th Annual International Symposium on Computer Architecture (ISCA)},
  pages={680--692},
  year={2020},
  organization={IEEE}
}

@inproceedings{lei2023put,
  title={Put Your Memory in Order: Efficient Domain-based Memory Isolation for {WASM} Applications},
  author={Lei, Hanwen and Zhang, Ziqi and Zhang, Shaokun and Jiang, Peng and Zhong, Zhineng and He, Ningyu and Li, Ding and Guo, Yao and Chen, Xiangqun},
  booktitle={Proceedings of the 2023 ACM SIGSAC Conference on Computer and Communications Security},
  pages={904--918},
  year={2023}
}

@inproceedings{lefeuvre2022flexos,
  title={{FlexOS}: towards flexible {OS} isolation},
  author={Lefeuvre, Hugo and B{\u{a}}doiu, Vlad-Andrei and Jung, Alexander and Teodorescu, Stefan Lucian and Rauch, Sebastian and Huici, Felipe and Raiciu, Costin and Olivier, Pierre},
  booktitle={Proceedings of the 27th ACM International Conference on Architectural Support for Programming Languages and Operating Systems},
  pages={467--482},
  year={2022}
}

@inproceedings{lu2023moat,
  title={{MOAT}: Towards Safe {BPF} Kernel Extension},
  author={Lu, Hongyi and Wang, Shuai and Wu, Yechang and He, Wanning and Zhang, Fengwei},
  booktitle={33rd USENIX Security Symposium (USENIX Security 24)},
  pages={1153--1170},
  year={2024}
}

@inproceedings{lin2024fast,
  title={Fast core scheduling with userspace process abstraction},
  author={Lin, Jiazhen and Chen, Youmin and Gao, Shiwei and Lu, Youyou},
  booktitle={Proceedings of the ACM SIGOPS 30th Symposium on Operating Systems Principles},
  pages={280--295},
  year={2024}
}

@inproceedings{maar2024beyond,
  title={Beyond the Edges of Kernel Control-Flow Hijacking Protection with {HEK-CFI}},
  author={Maar, Lukas and Nasahl, Pascal and Mangard, Stefan},
  booktitle={Proceedings of the 19th ACM Asia Conference on Computer and Communications Security},
  pages={1214--1230},
  year={2024}
}

@inproceedings{song2024interstellar,
  title={Interstellar: Fully Partitioned and Efficient Security Monitoring Hardware Near a Processor Core for Protecting Systems against Attacks on Privileged Software},
  author={Song, YongHo and Woo, Byeongsu and Han, Youngkwang and Kang, Brent ByungHoon},
  booktitle={Proceedings of the 2024 on ACM SIGSAC Conference on Computer and Communications Security},
  pages={198--212},
  year={2024}
}

@inproceedings{ahmad2023veil,
  title={Veil: A protected services framework for confidential virtual machines},
  author={Ahmad, Adil and Ou, Botong and Liu, Congyu and Zhang, Xiaokuan and Fonseca, Pedro},
  booktitle={Proceedings of the 28th ACM International Conference on Architectural Support for Programming Languages and Operating Systems, Volume 4},
  pages={378--393},
  year={2023}
}

@article{liu2025more,
  title={More Granular, Less Trust: Enforcing Intra-Process Isolation With {Arm} {CCA} in an Untrusted Management Environment},
  author={Liu, Shiqi and Jiang, Zhouqi and Wang, Jie and Zhou, Wei and Sun, Kun and Chen, Zhaohui and Xie, Yulai},
  journal={IEEE Transactions on Information Forensics and Security},
  volume={20},
  pages={12507--12522},
  year={2025},
  publisher={IEEE}
}

@inproceedings{lefeuvre2023assessing,
  title={Assessing the impact of interface vulnerabilities in compartmentalized software},
  author={Lefeuvre, Hugo and B{\u{a}}doiu, Vlad-Andrei and Chen, Yi and Huici, Felipe and Dautenhahn, Nathan and Olivier, Pierre},
  booktitle={Network and Distributed System Security Symposium (NDSS)},
  year={2023}
}

@inproceedings{sartakov2022cap,
  title={{CAP-VMs}: {Capability-Based} Isolation and Sharing in the Cloud},
  author={Sartakov, Vasily A and Vilanova, Llu{\'\i}s and Eyers, David and Shinagawa, Takahiro and Pietzuch, Peter},
  booktitle={16th USENIX Symposium on Operating Systems Design and Implementation (OSDI 22)},
  pages={597--612},
  year={2022}
}

@inproceedings{watson2015cheri,
  title={{CHERI}: A hybrid capability-system architecture for scalable software compartmentalization},
  author={Watson, Robert NM and Woodruff, Jonathan and Neumann, Peter G and Moore, Simon W and Anderson, Jonathan and Chisnall, David and Dave, Nirav and Davis, Brooks and Gudka, Khilan and Laurie, Ben and others},
  booktitle={2015 IEEE Symposium on Security and Privacy},
  pages={20--37},
  year={2015},
  organization={IEEE}
}

\vfill

\begin{IEEEbiography} [{\includegraphics[width=1in,height=1.25in,clip,keepaspectratio]{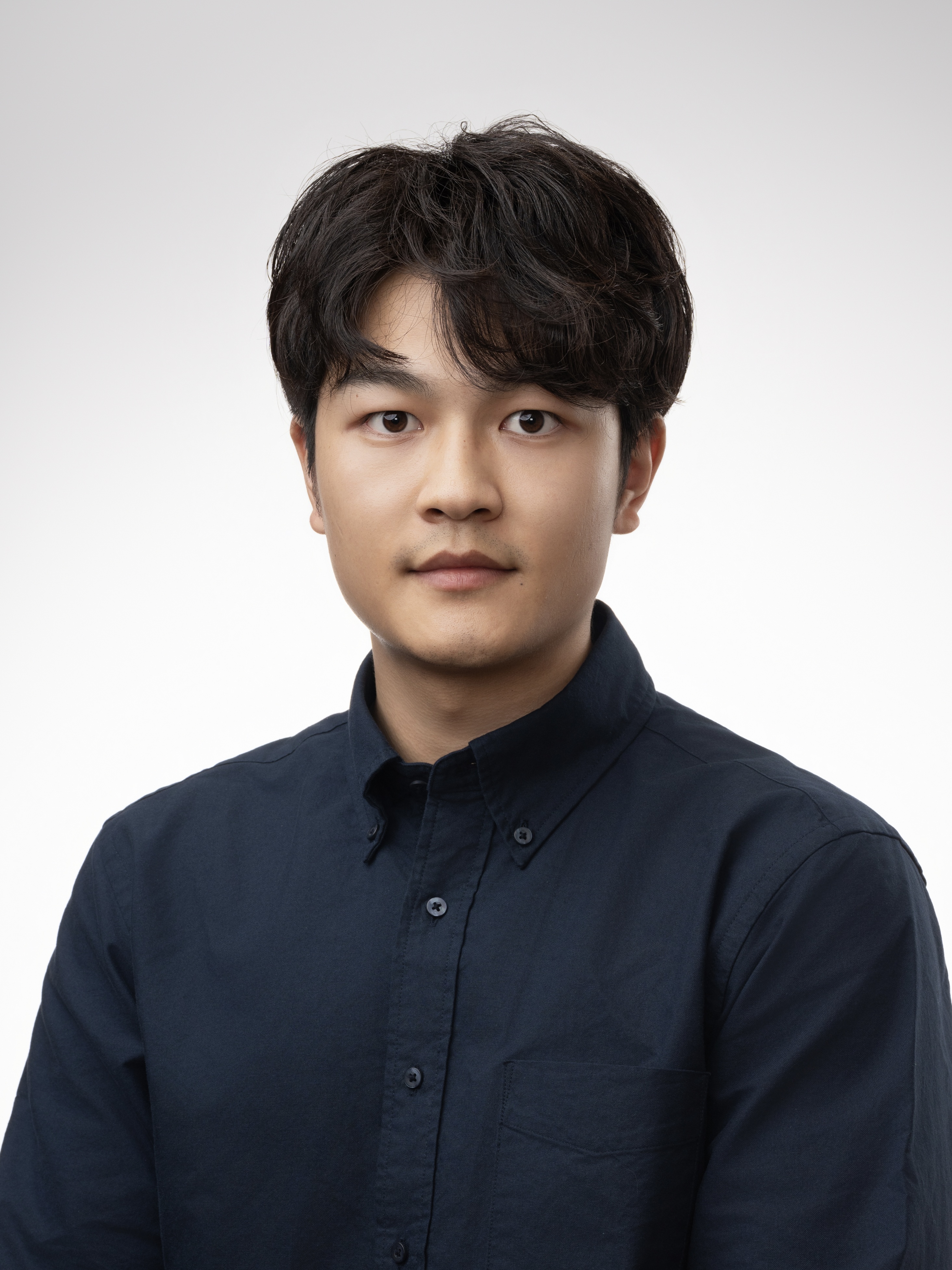}}]
{Shiqi Liu} 
received the B.S. degree in Cyber Science and Engineering from the University of International Relations, Beijing, China, in 2022, and the M.S. degree in Cyber Science and Engineering from Huazhong University of Science and Technology, Wuhan, China, in 2025. His research interests include hardware–software co-design and verification.
\end{IEEEbiography}

\begin{IEEEbiography} [{\includegraphics[width=1in,height=1.25in,clip,keepaspectratio]{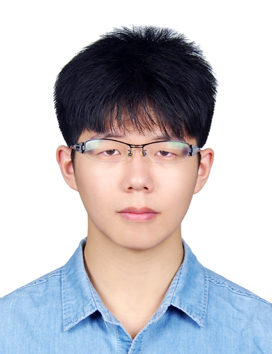}}]
{Yongpeng Gao}
received his B.S. and M.S. degrees from the School of Cyber Science and Engineering, Huazhong University of Science and Technology, Wuhan, China, in 2023 and 2026, respectively. His research interests include system security and software security.
\end{IEEEbiography} 

\begin{IEEEbiography} [{\includegraphics[width=1in,height=1.25in,clip,keepaspectratio]{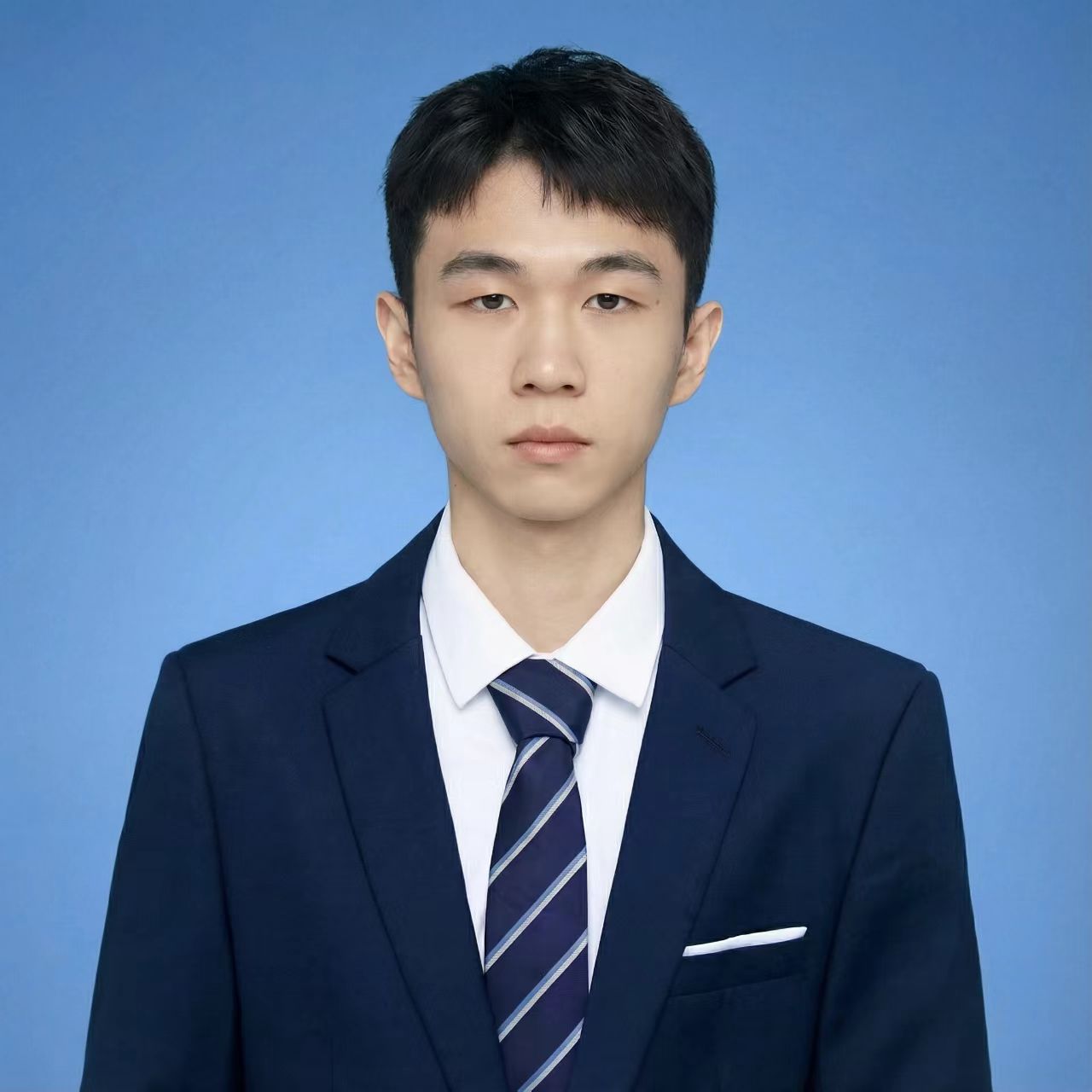}}]
{Mingyang Zhang}
received his B.S. degree in the School of Cyber Science and Engineering, Huazhong University of Science and Technology, Wuhan, China, in 2024. He is currently working toward an M.S. degree at the same institution. His research interests include system security and software security.
\end{IEEEbiography} 

\begin{IEEEbiography} [{\includegraphics[width=1in,height=1.25in,clip,keepaspectratio]{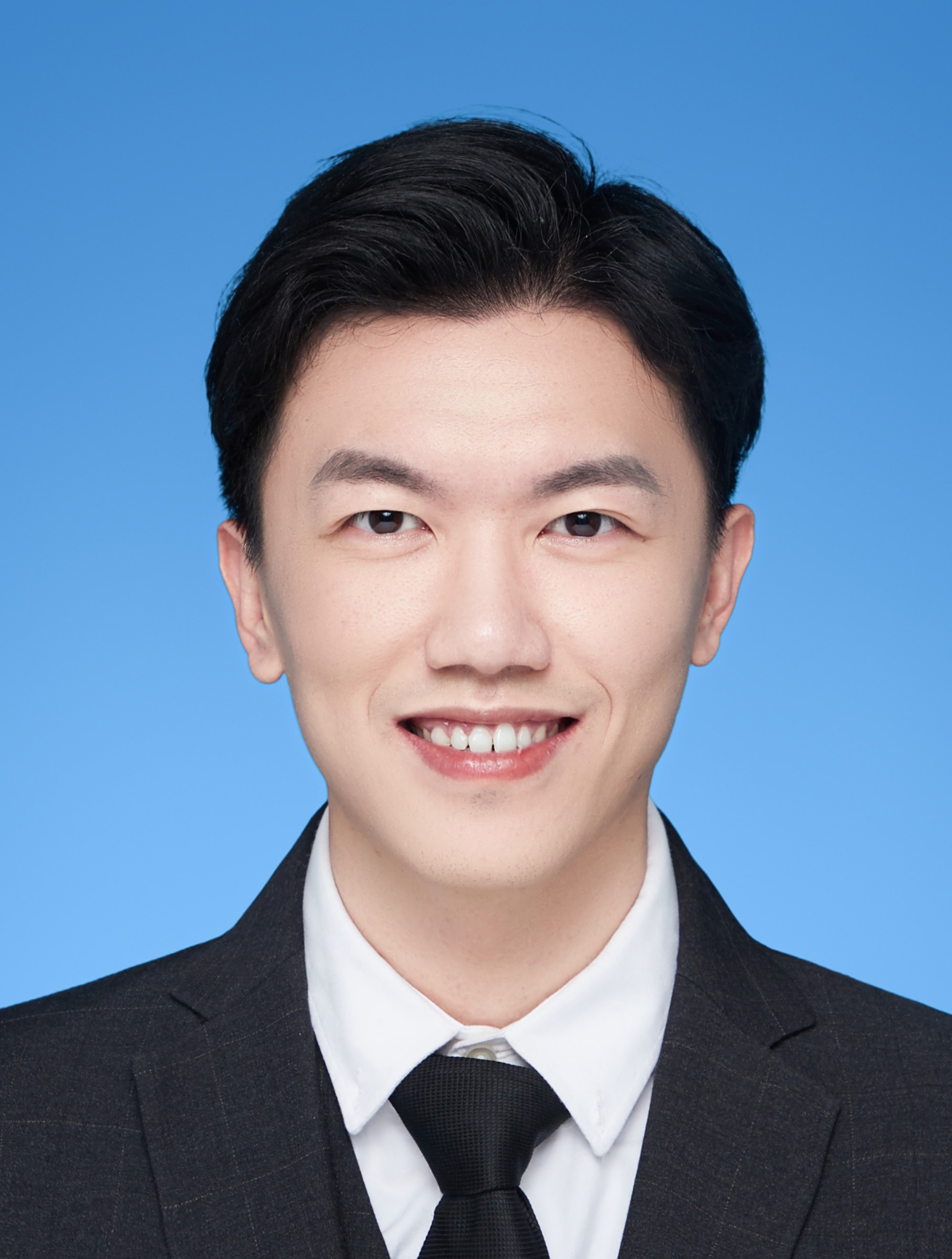}}]
{Jie Wang}
received his Ph.D. degree from the University of Chinese Academy of Sciences (UCAS) in 2021. He was a visiting scholar at George Mason University and is currently an Associate Professor at the School of Cyber Science and Engineering, Huazhong University of Science and Technology. His research interests include system security, trusted computing, and trusted AI accelerators.
\end{IEEEbiography} 

\end{document}